\documentclass[nofootinbib,preprintnumbers,amsmath,amssymb,aps,prc,twocolumn, showkeys,floatfix,superscriptaddress]{revtex4-1}

\usepackage{amsmath,graphics,color}
\usepackage{graphicx} 
\usepackage{booktabs, tabularx}
\usepackage{multirow,array}
\usepackage{graphics}
\usepackage{hyperref}
\usepackage{bm}
\usepackage{calligra}
\usepackage[T1]{fontenc}
\usepackage{amssymb}
\usepackage{amsfonts}
\usepackage{mathrsfs}  
\usepackage{dsfont}
\usepackage{footmisc}
\usepackage[normalem]{ulem}
\usepackage{xcolor}
\usepackage{caption}
\usepackage{float}
\usepackage{orcidlink}
\usepackage{subcaption}
\usepackage{ragged2e} 
\DeclareCaptionFormat{myformat}{\justifying#1#2#3}
\graphicspath{{figures/}}

\begin{document}

\title{Probing radial flow via the differential flow coefficient $v_0(p_T)$ using CMS open data}

\author{Allan E. F. G. Ferreira {\sc }\orcidlink{0009-0003-5143-4212}}
\email{allan.godoi@unesp.br}
\affiliation{
 Instituto de F\'isica Te\'orica (IFT), Universidade Estadual Paulista J\'ulio de Mesquita Filho (UNESP), R. Dr. Bento Teobaldo Ferraz, 271 - V\'arzea da Barra Funda, S\~ao Paulo - SP, Brazil, 01140-070
}

\author{Thiago S. Domingues {\sc }\orcidlink{0000-0001-6756-8600}}
\email{thiago.siqueira.domingues@usp.br}
\affiliation{
 Instituto de F\'isica, Universidade de S\~ao Paulo (USP), R. do Mat\~ao, 1371, S\~ao Paulo, Brazil, 05508-090
}%

\author{Matthew Luzum {\sc }\orcidlink{0000-0002-0367-7055}}
\email{mluzum@usp.br}
\affiliation{
 Instituto de F\'isica, Universidade de S\~ao Paulo (USP), R. do Mat\~ao, 1371, S\~ao Paulo, Brazil, 05508-090
}%

\author{Cesar A. Bernardes {\sc }\orcidlink{0000-0001-5790-9563}}
\email{cesar.bernardes@unesp.br}
\affiliation{
 N\'ucleo de Computa\c{c}\~ao Cient\'ifica (NCC), Universidade Estadual Paulista J\'ulio de Mesquita Filho (UNESP), R. Dr. Bento Teobaldo Ferraz, 271 - V\'arzea da Barra Funda, S\~ao Paulo - SP, Brazil, 01140-070 
}
\date{\today}
\begin{abstract}
We present a comprehensive study of the differential radial flow coefficient $v_0(p_T)$ and its $p_T$-integrated counterpart $v_0$ in Pb+Pb collisions at $\sqrt{s_{NN}} = 2.76~\mathrm{TeV}$, using CMS open data for mid-peripheral centrality classes of 50--60\% and 60--70\%. The measured $v_0(p_T)$ distributions are compared with ATLAS measurements at $\sqrt{s_{NN}} = 5.02~\mathrm{TeV}$, revealing a clear energy dependence in the overall magnitude of radial flow fluctuations. Notably, the scaled ratio $v_0(p_T)/v_0$ exhibits an approximate universality that is largely independent of both collision energy and centrality at low $p_T$, persisting up to $p_T\sim3~\mathrm{GeV}$, suggesting that the integrated $v_0$ encodes most of the energy-dependent information. We compare the data with the \textsc{Jetscape} multi-stage hybrid model, using Gaussian process emulators trained on a principal component analysis of the model output to explore the full 17-dimensional parameter space. A global variance-based sensitivity analysis using Sobol indices shows that, for the Grad viscous correction, the nucleon fluctuation parameter $\sigma_k$ is the dominant source of variance for $v_0$, $v_0(p_T)$, and $v_0(p_T)/v_0$, followed by the free-streaming parameters $\tau_R$ and $\alpha$ and the nucleon width $w$. The sensitivity to the bulk viscosity peak $(\zeta/s)_{\max}$ is weak for $v_0$ but increases for $v_0(p_T)$ and for the scaled ratio. Emulator predictions, calibrated on $p_T$-integrated observables and on scaled spectra, describe the data reasonably well at low and intermediate $p_T$. Parameters calibrated at $\sqrt{s_{NN}} = 2.76~\mathrm{TeV}$ also describe the ATLAS $v_0(p_T)/v_0$ data at $5.02~\mathrm{TeV}$, supporting a near universal momentum dependence of radial flow fluctuations across collision energies.  
\end{abstract}
\maketitle
\section{Introduction}
\label{sec:intro}
Ultra-relativistic heavy-ion collisions performed at facilities such as the \textit{Large Hadron Collider} (LHC) produce a strongly-interacting state of deconfined quarks and gluons known as the quark-gluon plasma (QGP). One of the most striking signatures of this medium is the emergence of collective flow, which reflects the hydrodynamic response of the system to the initial-state geometry and fluctuations. The collective expansion of the QGP can be decomposed into anisotropic and isotropic components. The anisotropic component is commonly characterized through the Fourier coefficients $v_n$ of the azimuthal particle distribution, while the isotropic component is associated with radial flow, corresponding to the collective outward expansion of the medium driven by pressure gradients. This radial expansion produces a characteristic hardening of the transverse momentum ($p_T$) spectra and generates the well-known mass ordering pattern observed in identified particle spectra and flow observables. 

A new transverse momentum differential observable for radial flow fluctuations, denoted by $v_0(p_T)$, has recently been proposed and extensively investigated~\cite{Gardim:2019iah,Schenke:2020uqq,Parida:2024ckk,Du:2025dpu,Saha:2025nyu,Zhu:2026vql,Bhatta:2025oyp,Wan:2025rzg,Sambataro:2025pop,Du:2025hrz,Agarwala:2025bxs,Parida:2026xbj,Jia:2025rab, Zhu:2026vql, Wan:2025rzg, Thakur:2026few, Sambataro:2025pop, Du:2025hrz}. The observable $v_0(p_T)$ measures the covariance between event-by-event fluctuations of the particle yield in a given $p_T$ interval and fluctuations of the event mean transverse momentum $[p_T]$, providing a direct probe of radial flow fluctuations. Formally, it is defined as~\cite{Schenke:2020uqq, Du:2025dpu}, 
\begin{equation}
\label{eq:radial_flow_coeff}
v_0(p_T)=
\frac{\langle \delta n(p_T)\,\delta[p_T]\rangle}
{\langle n(p_T)\rangle \sigma[p_T]},
\end{equation}
where $n(p_T)$ denotes the event-wise particle yield in a given $p_T$ interval, $[p_T]$ is the event mean transverse momentum, $\delta x = x-\langle x\rangle$ represents fluctuations relative to the ensemble average, and $\sigma[p_T]$ is the standard deviation of $[p_T]$. This definition is analogous to a Pearson correlation coefficient and quantifies how fluctuations of the spectral shape are correlated with fluctuations of the average transverse momentum. Furthermore, we can recognize the $p_T$-integrated counterpart of $v_0(p_T)$ as $v_0 \equiv \sigma[p_T]/\langle p_T \rangle$, since $\langle p_T \rangle = \langle [p_T] \rangle$.   

Recent theoretical and experimental studies have demonstrated that $v_0(p_T)$ exhibits several characteristic signatures of collective behavior. In particular, measurements by the ATLAS~\cite{ATLAS:2025ztg} and ALICE~\cite{ALICE:2025iud} collaborations in Pb+Pb collisions at $\sqrt{s_{NN}} = 5.02~\mathrm{TeV}$ show that $v_0(p_T)$ is dominated by long-range correlations and displays weak dependence on pseudorapidity gaps, indicating that the observable is primarily sensitive to global collective dynamics rather than short-range nonflow contributions~\cite{ALICE:2023tej,ALICE:2024apz,ATLAS:2024jvf,ATLAS:2025ztg,ALICE:2025iud, CMS:2026oug, Wang:2026tvi}. Furthermore, the normalized shape $v_0(p_T)/v_0$ exhibits an approximately universal behavior across collision centralities, closely resembling the scaling properties observed in anisotropic flow measurements~\cite{Parida:2026xbj}. Another remarkable feature is the sign change of $v_0(p_T)$ near the average transverse momentum of the spectra $\langle p_T \rangle$, which naturally emerges from radial flow fluctuations: events with stronger radial flow produce harder spectra, yielding positive correlations at large $p_T$ and negative correlations at low $p_T$~\cite{Schenke:2020uqq}. These properties strongly support the interpretation of $v_0(p_T)$ as a genuine collective observable associated with isotropic hydrodynamic expansion. While anisotropic flow observables $(v_n , n = 1, 2, \dots)$ mainly constrain shear viscous effects~\cite{Song:2009rh, Qian:2016fpi}, radial flow fluctuations can probe the isotropic response of the medium and therefore provide complementary sensitivity to bulk viscous dynamics. Recent hydrodynamic calculations further suggest that the scaled observable $v_0(p_T)/v_0$ is particularly sensitive to the bulk viscosity of the QGP~\cite{Schenke:2020uqq,ATLAS:2025ztg}, 
and consequently, this observable has emerged as a promising tool for constraining QGP transport coefficients and for studying the interplay between initial-state fluctuations and the isotropic hydrodynamic response~\cite{Du:2025dpu}.

In this work, we perform a systematic investigation of the radial flow observable $v_0(p_T)$ for charged particles using CMS public data~\cite{cms_open_data} for Pb+Pb collisions at $\sqrt{s_{NN}}=2.76~\mathrm{TeV}$ and compare our results with measurements from the ATLAS Collaboration at $\sqrt{s_{NN}}=5.02~\mathrm{TeV}$ for charged particles~\cite{ATLAS:2025ztg}. Specifically, we study the centrality dependence, scaling properties, and transverse momentum structure of $v_0(p_T)$ exploring its sensitivity to the underlying medium properties through a global sensitivity analysis over the parameter space of a state-of-the-art hybrid model of heavy-ion collisions. Motivated by the recent evidence for the collective nature of radial flow fluctuations~\cite{Du:2025dpu}, we employ Gaussian Process (GP) emulators to scan the entire parameter space of state-of-the-art hybrid models~\cite{JETSCAPE:2020mzn} against the measured observables. This paper is organized as follows. In Section~\ref{sec:analysis_setup}, we describe the analysis framework employed in this study,  where in Subsection~\ref{subsec:exp_and_th} we present the experimental data sets and analysis methodology, and in Subsection~\ref{subsec:emulator_v0pt} we explain in detail the construction and validation of the emulators used to predict the radial flow observable over the parameter space, while Subsection~\ref{subsec:gsa} describes the global sensitivity analysis employed to identify the model parameters most relevant to $v_0(p_T)$ and its counterpart $v_0$. In the results Section~\ref{sec:results}, we start exploring the collective properties of $v_0(p_T)$ using experimental data from CMS in Subsection~\ref{subsec:exp_v0pt}, followed by a full parameter space scan in Subsec.~\ref{subsec:emulators} using trained emulators, concluding this section by scaling $v_0$ and comparing this scaled observable with experimental data in Subsection~\ref{subsec:scaled_v0pt}. Our conclusions are presented in Section~\ref{sec:conclusions}. Additional information regarding the model calculations, Principal Component Analysis (PCA), and emulator validation is provided in the Appendix~\ref{appendix:appendix_dpts_and_pca} and~\ref{appendix:emu_validation}.
\section{Analysis setup}
\label{sec:analysis_setup}
\subsection{Experimental data and analysis method} 
\label{subsec:exp_and_th}
The data analysis is carried out using a publicly-available dataset from the CMS open data portal, corresponding to Pb+Pb collisions at $\sqrt{s_{NN}}=2.76~\mathrm{TeV}$ collected during the 2011 run~\cite{cms_open_data}, restricted to peripheral collisions in the 50--70\% centrality class~\footnote{The present analysis is restricted to events with centrality 50-70\%, as this dataset, provided by the CMS Collaboration, does not include more central collision classes.}. Events and tracks are selected following the CMS prescriptions for the dataset~\cite{CMS:2013bza}. Track selection criteria include $0.5 \leq p_T \leq 10.0$ GeV and pseudorapidity $|\eta|<2.4$.

The quantities of interest are calculated as: $[p_T]=\sum_i w_ip_T/\sum_i w_i$ and $n(p_T)=\sum_i^{p_T}w_i/\sum_iw_i$, where $\sum_i^{p_T}$ denotes the sum over all tracks in a small $p_T$ bin and $w_i$ are the detector correction weights. These corrections are derived using full Monte Carlo simulations of Pb+Pb collisions with the \textsc{Hydjet}~\cite{Lokhtin:2005px} event generator and \textsc{Geant4}~\cite{AGOSTINELLI2003250}, by comparing charged particles in the generator level with reconstructed tracks \cite{CMS:2012zex, CMS:2013bza}.

In order to suppress non-flow correlations, we used a 2-subevent method, splitting the dataset into two $\eta$ ranges $-2.4 \leq \eta_A \leq -\eta_{gap}/2$ and $\eta_{gap}/2 \leq \eta_B \leq 2.4$, separated by a zero-centered interval of length $\eta_{gap}$. We use different $\eta_{gap}$ in the analysis to investigate long-range correlations \cite{ATLAS:2025ztg, atlas-qm25, atlas-is25}.

The radial flow coefficient $v_0(p_T)$ can be obtained using a factorization assumption, dividing the normalized covariance $\langle \delta n(p_T) \delta [p_T] \rangle / \langle n(p_T) \rangle \langle p_T \rangle$ by the $p_T$-integrated observable $v_0$, thus recovering Eq.  \ref{eq:radial_flow_coeff}.
\begin{equation}
    v_0(p_T) = \frac{1}{v_0}\frac{\langle \delta n(p_T) \delta [p_T] \rangle}{\langle n(p_T) \rangle \langle p_T \rangle}.
\label{eq:v0pt_definition}    
\end{equation}
We validate the factorization assumption by constructing the observables evaluating $\delta[p_T]$ and $\langle p_T \rangle$ in different $p_T$ reference ($p_T^{ref}$) ranges; for $\delta n(p_T)$ and $\langle n( p_T)\rangle$, we use tracks in each analysis bin within $0.5 < p_T < 10.0 \text{ GeV}$. 

Theoretical predictions for the observables presented in this paper are obtained with the \textsc{Jetscape} framework~\cite{JETSCAPE:2020mzn,JETSCAPE:2020shq, Domingues:2026mis}, adopting an identical binning scheme to that applied to the experimental data.
\subsection{Principal Component Analysis  and Gaussian Process Regressors}
\label{subsec:emulator_v0pt}
To enable fast and accurate predictions of the integrated radial flow coefficient, the differential radial flow coefficient, and their ratio across the full parameter space of the \textsc{Jetscape} hybrid heavy-ion collision model, we construct GP emulators for the principal component representation obtained from the PCA described in Appendix Sec.~\ref{appendix:appendix_dpts_and_pca}. This surrogate modeling approach allows us to interpolate the high-dimensional model output with quantified uncertainties and has been extensively used in modern Bayesian parameter estimation of relativistic heavy-ion collisions~\cite{Domingues:2026mis, JETSCAPE:2020mzn, Roch:2024xhh, Jahan:2025uye, Auvinen:2026dyl}. 

The emulator inputs consist of the full set of $N_{\text{mp}}$ model parameters defining the initial conditions, transport coefficients, and particlization of the hybrid simulation framework. See Ref.~\cite{Domingues:2026mis} for the full Table of parameters. These parameters include the initial entropy normalization, nucleon substructure properties, relaxation time scales, shear and bulk viscosity temperature parameterizations, and the switching temperature. The design points are sampled within predefined uniform prior distributions over a range using a space-filling design, resulting in $N_{\text{design}}$ training points in the $N_{\text{mp}}$-dimensional parameter space. Specifically, in this analysis, we have a 17-dimensional parameter space. Each design point is represented by the parameter vector
\begin{equation}
\boldsymbol{\theta}_i = (\theta_{i1}, \theta_{i2}, \ldots, \theta_{iN_{\text{mp}}}),
\label{eq:design_vector}
\end{equation}
which serves as the input to the emulator. 

Rather than emulating the full model output directly as a function of centrality and transverse momentum, we first perform a PCA on the observables, thereby reducing the dimensionality of the emulator target space.  Independent emulators are then trained for the first $N_{\mathrm{PC}}$ principal component coefficients, 
\begin{equation}
\mathbf{z}_i = (z_{i1}, z_{i2}, \ldots, z_{iN_{\mathrm{PC}}}),
\label{eq:pca_scores}
\end{equation}
where each principal component represents an orthogonal mode of variation of the observable space. This decomposition substantially reduces the complexity of the emulation problem while preserving the dominant physical information encoded in the original model calculations. 

In practice, only a small number of principal components are required to accurately reconstruct the observables. We retain the three principal components for $v_0$, which already capture more than $95\%$ of the total variance of the full dataset and the first 9 principal components for $v_0(p_T)$, retaining more than $90\%$ of the total variance. See Appendix~\ref{appendix:appendix_dpts_and_pca} for further details. For each principal component $k$, we train an independent GP regressor,
For each principal component k, we train an independent GP regressor,
\begin{equation}
z_k(\boldsymbol{\theta})
\sim
\mathcal{GP}
\left(
0,
K_k(\boldsymbol{\theta},\boldsymbol{\theta}')
\right),
\label{eq:pca_scores_emu}
\end{equation}
with covariance kernel
\begin{equation}
K_k = C \times \mathrm{RBF}(\boldsymbol{\ell}) + \mathrm{WhiteNoise}.
\label{eq:GP_kernel}
\end{equation}
The constant kernel C controls the overall covariance amplitude, while the radial basis function (RBF) kernel introduces smooth correlations in parameter space through a characteristic length-scale vector,
\begin{equation}
\boldsymbol{\ell} = 
(\ell_1,\ell_2,\ldots,\ell_{N_{\mathrm{mp}}}),
\label{eq:ch_length_scale}
\end{equation}
with one independent length scale associated with each model parameter dimension. This anisotropic kernel structure allows the emulator to adaptively learn the relative sensitivity of each principal component to variations in different physical parameters. The characteristic length scales are initialized using the parameter ranges of the design and optimized during training within physically motivated bounds. The WhiteNoise component accounts for residual statistical fluctuations from the finite event-by-event simulations, imperfect PCA reconstruction, and numerical uncertainties in the hybrid model calculations. The GP hyperparameters are determined by maximizing the log-marginal likelihood of the training data,
\begin{equation}
\log \mathcal{L} =
-\frac{1}{2}\mathbf{z}^T K^{-1}\mathbf{z}
- \frac{1}{2}\log|K| - \frac{N}{2}\log(2\pi),
\label{eq:likelihood_optm}
\end{equation}
where $K$ denotes the covariance matrix evaluated on the training design points.
To avoid convergence to local extrema, the optimization is repeated using multiple random restarts of the optimizer. The emulator implementation employs \href{https://scikit-learn.org/stable/modules/gaussian_process.html}{Scikit-learn}~\cite{scikit-learn} python library, following methodologies commonly adopted in Bayesian analyses of heavy-ion collisions~\cite{Bernhard:2019bmu,JETSCAPE:2020mzn,Nijs:2020roc,Nijs:2020ors,Parkkila:2021tqq,Mantysaari:2022ffw,Heffernan:2023gye}.

The emulators are trained independently for each viscous correction prescription (Grad~\cite{Grad:1949zza, Dusling:2009df, McNelis:2019auj, Teaney:2003kp, Monnai:2009ad, Dusling:2011fd, Denicol:2012cn}, CE~\cite{Anderson:1974nyl, Jaiswal:2014isa}, PTM~\cite{Pratt:2010jt, McNelis:2019auj}, and PTB~\cite{Bernhard:2016tnd,Pratt:2010jt}). For a given correction scheme, a separate set of GP emulators is constructed using the corresponding PCA-transformed training data. This strategy allows us to isolate the impact of viscous corrections on the structure of the radial flow observables while maintaining a consistent emulator architecture across all cases. 

The resulting emulator framework provides a smooth and fully differentiable surrogate model for the radial flow observables $v_0$, $v_0(p_T)$, and $v_0(p_T)/v_0$. Once trained, the emulators can be evaluated at negligible computational cost, enabling efficient uncertainty quantification, global sensitivity analysis, and Bayesian parameter inference over the full 17-dimensional model parameter space without requiring additional expensive event-by-event hybrid simulations.
\subsection{Global sensitivity analysis}
\label{subsec:gsa}
To quantify the relative importance of the model parameters on the observables, we perform a global sensitivity analysis (GSA) based on variance decomposition using Sobol indices~\cite{SOBOL2001271}. 
This approach identifies the dominant sources of output uncertainty while fully accounting for nonlinear effects and parameter interactions across the entire prior parameter space. 
Direct evaluation of Sobol indices using full event-by-event simulations is computationally prohibitive; we therefore use emulators defined in Sec.\ref{subsec:emulator_v0pt} as fast surrogates. The GSA is performed independently for each viscous-correction scheme considered in this study, using the corresponding trained emulator set.

\paragraph{Sobol decomposition.}
Let $Z = f(\boldsymbol{\theta})$ denote the emulator prediction for a given PC coefficient, where $\boldsymbol{\theta} = (\theta_1, \dots, \theta_{N_\text{mp}})$ is the vector of $N_\text{mp}$ model parameters sampled within their prior ranges. Assuming mutual independence of the parameters, the ANOVA-HDMR (high-dimensional model representation) functional decomposition~\cite{SOBOL2001271} gives
\begin{equation}
  f(\boldsymbol{\theta})
  = f_0
    + \sum_{r=1}^{N_\text{mp}} f_r(\theta_r)
    + \sum_{1 \le r < s \le N_\text{mp}} f_{rs}(\theta_r,\theta_s)
    + \cdots ,
  \label{eq:hdmr}
\end{equation}
where $f_0 = \mathbb{E}[Z]$ and every term has zero mean. Taking the variance of both sides and using the orthogonality of the summands yields
\begin{equation}
  \mathrm{Var}(Z)
  = \sum_{r=1}^{N_\text{mp}} V_r
    + \sum_{1 \le r < s \le N_\text{mp}} V_{rs}
    + \cdots ,
  \label{eq:sobol_decomp}
\end{equation}
where the partial variances are defined as
\begin{align}
  V_r    &= \mathrm{Var}\!\left(\mathbb{E}[Z \mid \theta_r]\right), 
  \label{eq:V_r}\\
  V_{rs} &= \mathrm{Var}\!\left(\mathbb{E}[Z \mid \theta_r,\theta_s]\right) - V_r - V_s ,
  \label{eq:V_rs}
\end{align}
and higher-order terms follow the same inclusion–exclusion pattern. The first-order Sobol index,
\paragraph{First- and total-order indices.}
The \emph{first-order} (main-effect) Sobol index,
\begin{equation}
  S_r = \frac{V_r}{\mathrm{Var}(Z)}
       = \frac{\mathrm{Var}\!\left(\mathbb{E}[Z \mid \theta_r]\right)}{\mathrm{Var}(Z)},
  \label{eq:S_r}
\end{equation}
measures the fractional contribution of $\theta_r$ alone to the
output variance.
The \emph{total-order} index,
\begin{equation}
  S_{T,r}
  = \frac{\mathbb{E}\!\left[\mathrm{Var}(Z \mid \boldsymbol{\theta}_{\sim r})\right]}
         {\mathrm{Var}(Z)}
  = 1 - \frac{\mathrm{Var}\!\left(\mathbb{E}[Z \mid \boldsymbol{\theta}_{\sim r}]\right)}
             {\mathrm{Var}(Z)},
  \label{eq:S_T}
\end{equation}
captures the total effect of $\theta_r$, including all interaction
terms in which it participates; here $\boldsymbol{\theta}_{\sim r} \equiv \{\theta_s : s \neq r\}$ denotes the set of all parameters except $\theta_r$. By construction,
\begin{equation}
  S_r \le S_{T,r}, \qquad
  \sum_{r=1}^{N_\text{mp}} S_r \le 1, \qquad
  \sum_{r=1}^{N_\text{mp}} S_{T,r} \ge 1,
  \label{eq:index_bounds}
\end{equation}
with equality in the first sum if the model is purely additive. In the following, we focus on the total-order indices, which provide a
robust ranking of parameter importance in the presence of strong
nonlinearities and cross-parameter correlations. 

\paragraph{Estimation procedure.}
The indices are estimated with the Saltelli sampling scheme~\cite{Saltelli2002}, as implemented in the \textsc{SALib} Python library~\cite{Iwanaga2022,Herman2017}. For each viscous-correction scheme we draw $N = 1024$ base samples from the prior, requiring $N(2N_\text{mp}+2)$ emulator evaluations per PC coefficient. The sampling bounds are the parameter ranges used for emulator training (see Table~I of Ref.~\cite{Domingues:2026mis}). We perform variance-based global sensitivity analysis using Sobol indices to identify the most influential model parameters for $v_0$, $v_0(p_T)$, and $v_0(p_T)/v_0$. For each observable, we report the total-order index $S_T$ and the first-order index $S_1$ for each model parameter, computed independently for each retained PC coefficient; the figures show results
for the leading PC, which carries the largest fraction of the observable
variance. The gap $S_T - S_1$ quantifies the contribution from
parameter interactions and nonlinear effects.
\section{RESULTS}
\label{sec:results}
\subsection{Probing collective properties of differential radial flow coefficient using CMS data}
\label{subsec:exp_v0pt}
\begin{figure*}[ht!]
    \centering
    \begin{subfigure}[b]{0.48\textwidth}
        \centering
        \includegraphics[width=\linewidth]{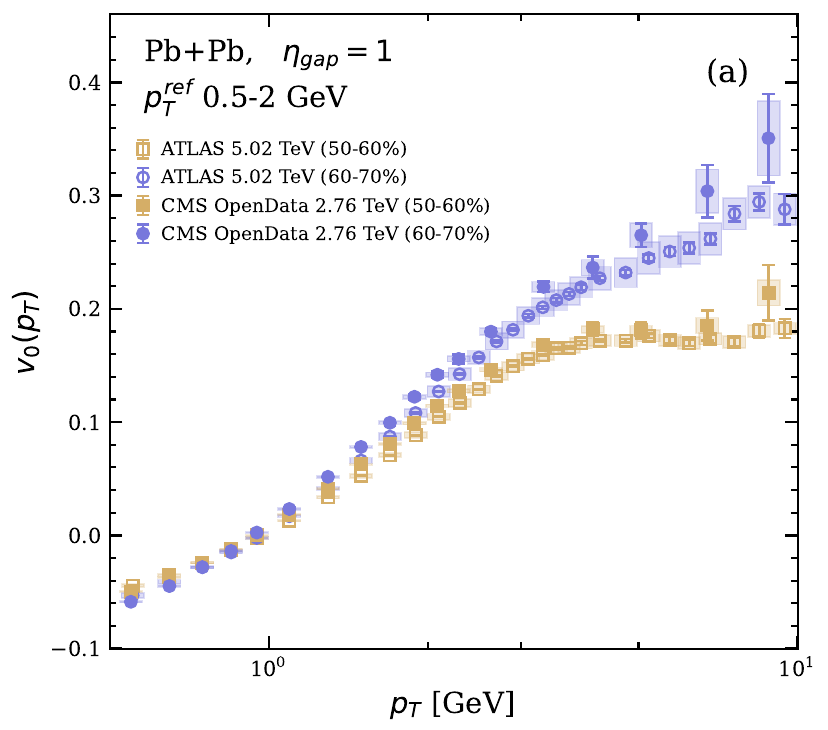}
    \end{subfigure}
    \begin{subfigure}[b]{0.48\textwidth}
        \centering
        \includegraphics[width=\linewidth]{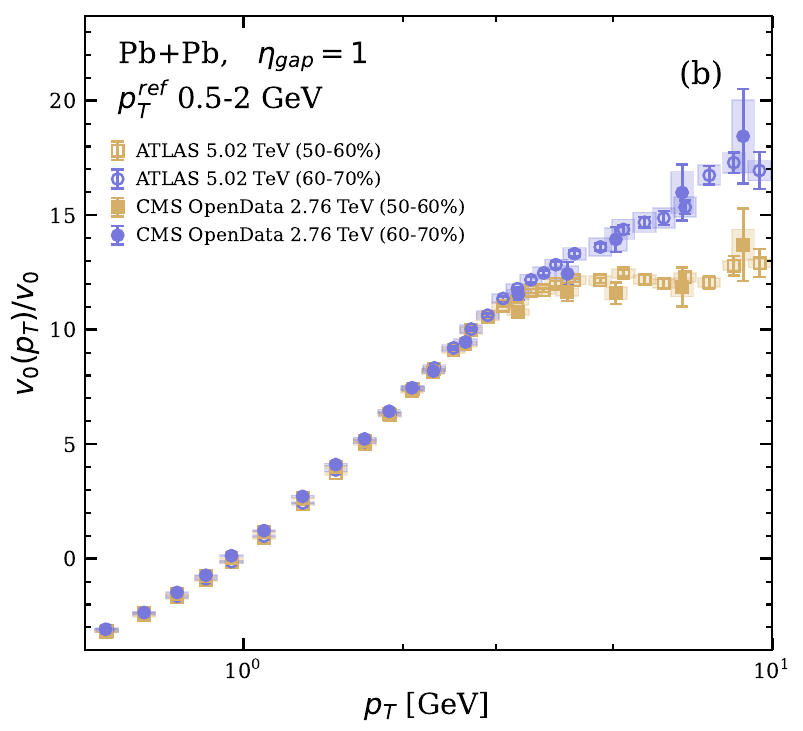}
    \end{subfigure}
    \caption{(a) Differential radial flow coefficient $v_0(p_T)$ and (b) scaled ratio $v_0(p_T)/v_0$ in the 50--60\% and 60--70\% centrality classes for Pb+Pb collisions at $\sqrt{s_{NN}}=2.76$ TeV (CMS open data) and $\sqrt{s_{NN}}=5.02$ TeV (ATLAS).  The analysis uses $\eta_{\mathrm{gap}}=1$ and $p_T^{\mathrm{ref}} \in [0.5,2]$ GeV.  Panel (a) shows a clear energy dependence in the overall magnitude of $v_0(p_T)$, with larger values at 2.76 TeV. Panel (b) demonstrates that scaling by the integrated $v_0$ largely removes this energy dependence, revealing an approximately universal momentum dependence that is nearly independent of centrality at low $p_T$ and remains similar up to $p_T \sim 3$ GeV.}
    \label{fig:plot-main}
\end{figure*}

We compare CMS Collaboration measurements for Pb+Pb at $\sqrt{s_{NN}}=2.76$ TeV~\cite{cms_open_data} with the published results from the ATLAS Collaboration for Pb+Pb collisions at $\sqrt{s_{NN}}=5.02$ TeV~\cite{ATLAS:2025ztg} to investigate variations due to the energy difference. The error boxes in the plots represent systematic uncertainties, while the error bars represent statistical uncertainties, which were estimated using a Poisson Bootstrap method \cite{bootstrap}. The systematic uncertainties stem from tracking selection, centrality calibration, and the position in the z-direction of the vertex. Effects from simultaneous collisions in the same or adjacent bunch crossing (pileup) are negligible for peripheral Pb+Pb collisions from the 2011 data-taking \cite{pileup}.
\begin{figure}
    \centering
    \includegraphics[width=\linewidth]{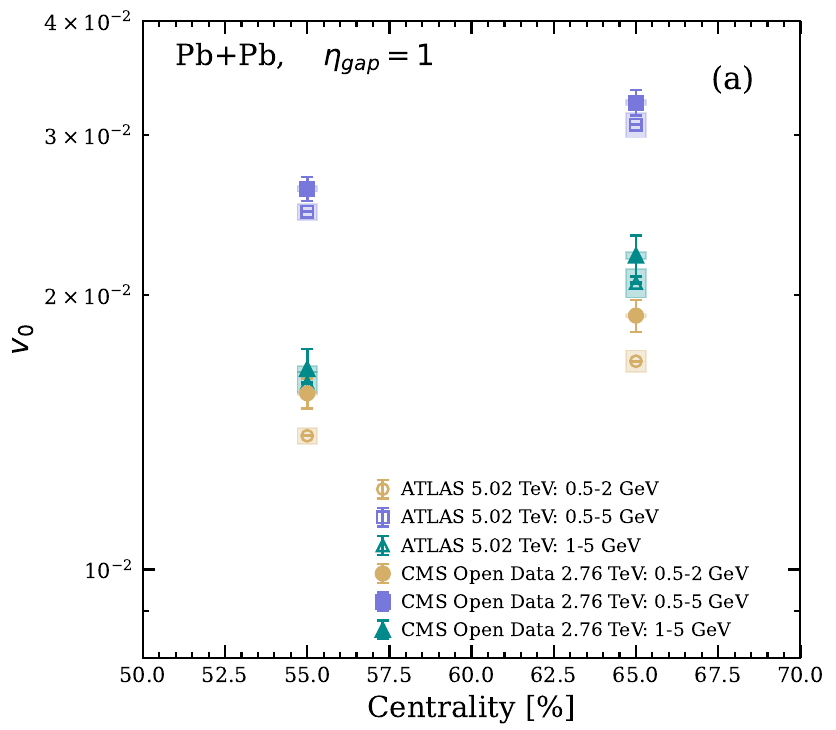}
    \includegraphics[width=\linewidth]{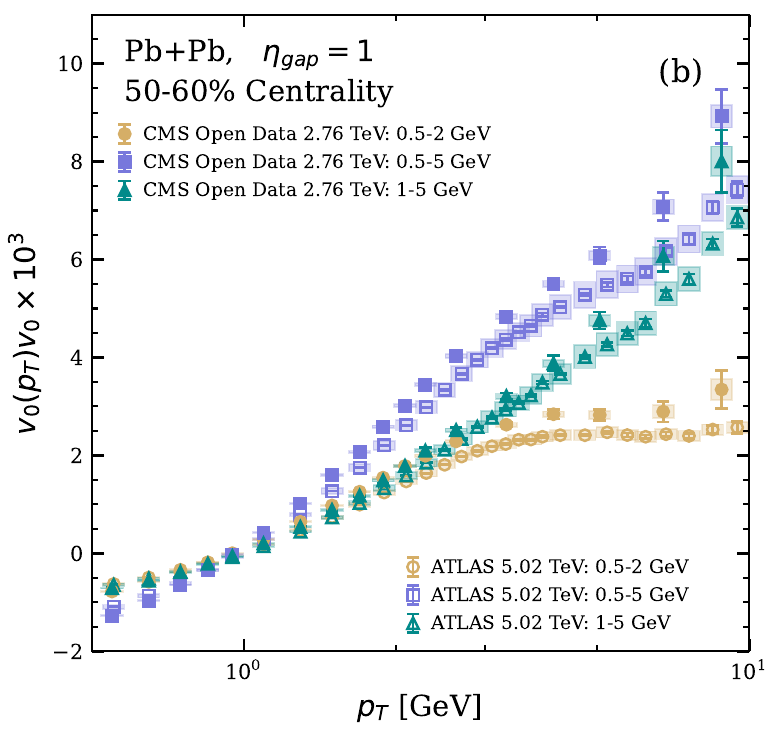}
    \caption{Integrated $v_0$ observable behavior in 50-60\% and 60-70\% centrality intervals (a) and normalized covariance depending on $p_T$ (b), both evaluated in 0.5-2, 0.5-5 and 1-5 GeV $p_T^{ref}$ ranges.}
    \label{fig:v0-cov}
\end{figure}
In Fig. \ref{fig:v0-cov}a and \ref{fig:v0-cov}b we observe respectively, the growth of $v_0$ as the collisions get more peripheral and the shape of the normalized covariance, highlighting the `seesaw' behavior when its value has a common zero-crossing point for all $p_T^{ref}$ ranges at $p_T \approx \langle p_T \rangle$ of the full track sample~\footnote{The $\langle p_T \rangle$ is expected to be similar for both collision energies (a bit higher for $5.02~\mathrm{TeV}$ due to enhanced radial push at higher collision energies~\cite{ALICE:2013wgn}), but in this analysis is even closer because we are measuring $\langle p_T \rangle$ in the $p_T$ range of 0.5--10 GeV}.
\begin{figure}
    \centering
    \includegraphics[width=\linewidth]{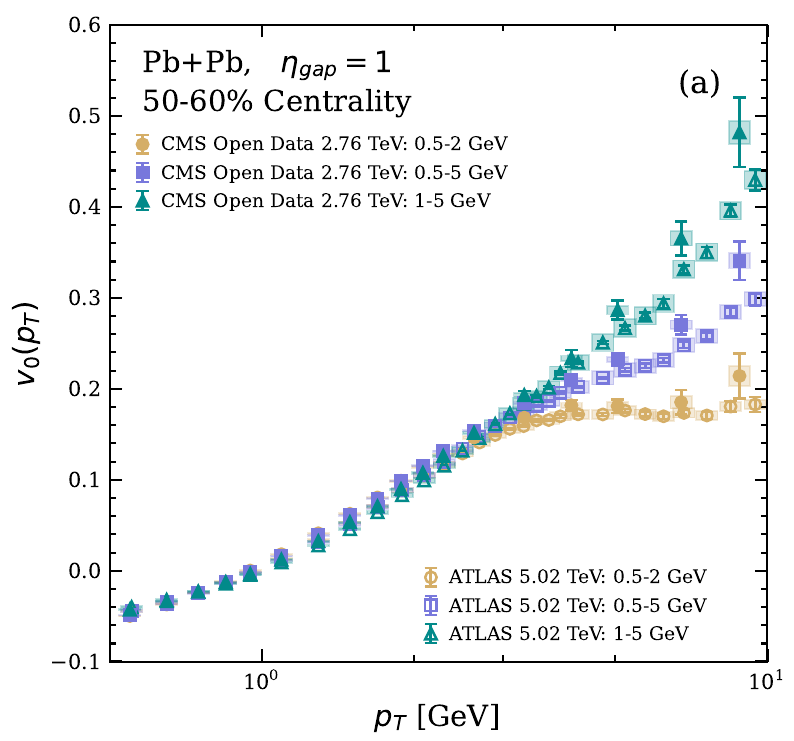}
    \includegraphics[width=\linewidth]{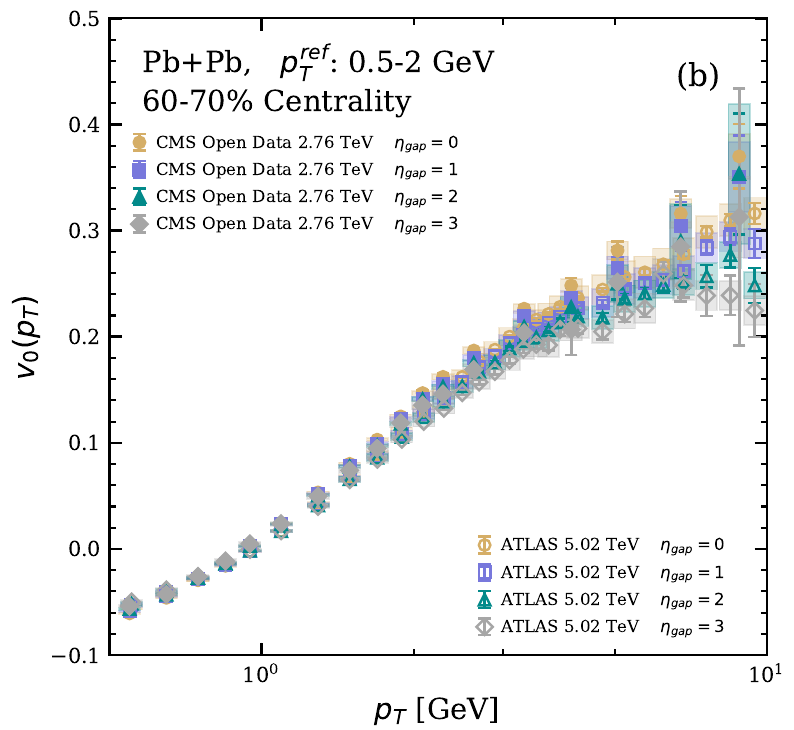}
    \caption{Differential radial flow coefficient $v_0(p_T)$ evaluated using different $p_T^{ref}$ ranges and fixed $\eta_{gap}=1$ in 50-60\% centrality (a) and different $\eta_{gap}$ and fixed $p_T^{ref}$ range [0.5-2] GeV in 60-70\% centrality (b).}
    \label{fig:v0pt-refs-eta}
\end{figure}
In Fig.~\ref{fig:v0pt-refs-eta}a, we observe that $v_0(p_T)$ exhibits negligible dependence on the choice of $p_T^{\mathrm{ref}}$ in the low-$p_T$ region (below around 3~GeV). This behavior supports the factorization assumption in this regime and is consistent with a description in terms of an effective one-body distribution~\cite{ATLAS:2024jvf}. The weak sensitivity to the reference interval indicates that the dominant correlations at low $p_T$ arise from collective dynamics rather than kinematic biases associated with the reference selection. A similar robustness is seen in Fig.~\ref{fig:v0pt-refs-eta}b when varying $\eta_{\mathrm{gap}}$. The observable remains largely unchanged across different pseudorapidity gaps, with only a mild dependence emerging at higher $p_T$. This weak sensitivity to detector acceptance suggests that $v_0(p_T)$ is dominated by long-range correlations, characteristic of collective flow phenomena rather than short-range nonflow effects. Adopting $\eta_{\mathrm{gap}}=1$ and $p_T^{\mathrm{ref}} \in [0.5,2]$ GeV as default parameters, Fig.~\ref{fig:plot-main}a shows $v_0(p_T)$ in the 50--60\% and 60--70\% centrality classes. A clear energy dependence is observed: the CMS data at $\sqrt{s_{NN}}=2.76$ TeV exhibit systematically larger values of $v_0(p_T)$ than the ATLAS data at $\sqrt{s_{NN}}=5.02$ TeV in the same centrality intervals. This difference reflects the overall magnitude of transverse-momentum fluctuations, which is sensitive to the collision energy and system dynamics. However, when considering the scaled ratio $v_0(p_T)/v_0$, shown in Fig.~\ref{fig:plot-main}b, the energy dependence is largely removed. In the low-$p_T$ region, the ratio is nearly independent of centrality and collision energy, indicating that the shape of the momentum dependence factorizes from the global fluctuation strength. Remarkably, this approximate universality persists up to $p_T \sim 3$ GeV, suggesting that the underlying mechanism governing the momentum dependence of radial flow fluctuations is largely energy independent within the explored range. The collapse of different energies onto a common curve after scaling by $v_0$ provides evidence for a near universal structure in the transverse-momentum correlations, with the integrated $v_0$ encoding most of the energy-dependent information. We note that in standard model-to-data analysis of heavy-ion collisions (e.g., Refs.~\cite{Bernhard:2019bmu, JETSCAPE:2020mzn, Domingues:2026mis}), radial flow fluctuations are incorporated via the integrated relative mean-$p_T$ variance, defined as $\frac{\delta p_T}{\langle p_T \rangle} = \frac{p_T - \langle p_T \rangle}{\langle p_T \rangle}$
, which $v_0$ is directly related to this quantity.
\subsection{Exploring the model parameter space}
\label{subsec:emulators}
\begin{figure}
    \centering
    \includegraphics[width=\linewidth]{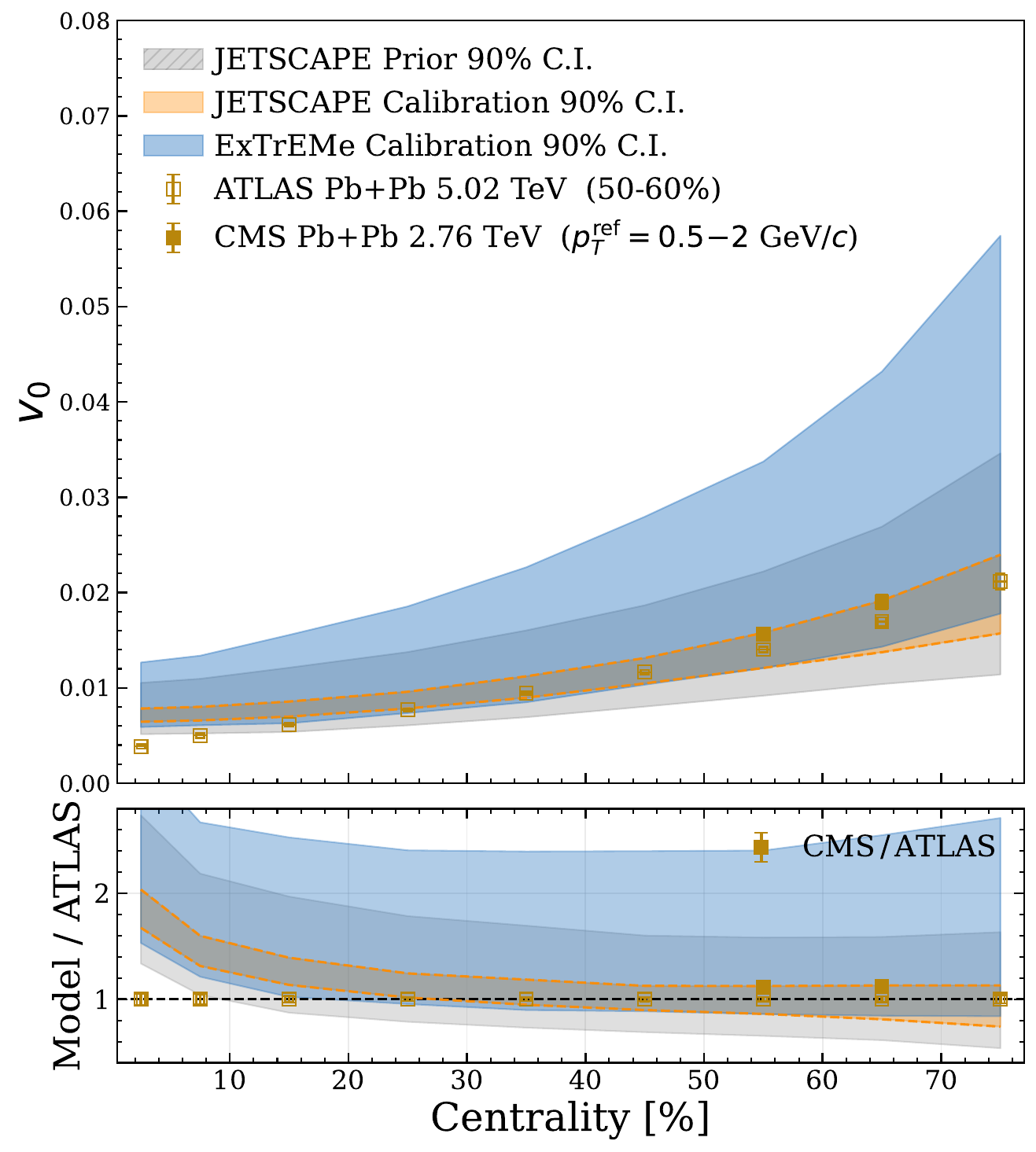}
    \caption{Global $v_0$ as a function of centrality for the Grad viscous correction. The gray band (labeled ``\textsc{Jetscape} Prior 90\% C.I.'') represents the 90\% credible interval obtained from the full prior parameter space from~\cite{JETSCAPE:2020mzn} for a Pb+Pb $\sqrt{s_{NN}} = 2.76$ parameters calibration. The colored bands correspond to posterior predictions from different calibration strategies: the blue band is obtained from a calibration including only $U(x_T)$ observables~\cite{Domingues:2026mis}, while the orange band corresponds to a calibration using only $p_T$-integrated observables within the \textsc{Jetscape} framework~\cite{JETSCAPE:2020mzn}. 
    We observe that the ATLAS Pb+Pb $\sqrt{s_{NN}} = 5.02$ TeV data are well reproduced by the posterior constrained by $p_T$-integrated observables (orange band), as expected. We included CMS Pb+Pb $\sqrt{s_{NN}} = 2.76$ TeV.}
    \label{fig:global_v0}
\end{figure}

In this section, we investigate the behavior of the model across its full parameter space. Such an exploration serves two main purposes. First, it provides a baseline characterization of the model’s intrinsic variability and predictive power under prior assumptions. Second, it establishes a reference against which posterior constraints can be quantitatively assessed. This approach follows the standard strategy in Bayesian analyses, where prior predictive distributions are examined before calibration to data in order to identify dominant sources of uncertainty and parameter sensitivities~\cite{Hu:2026ebv, Han_2022}. Using the same kinematic cuts as in the CMS and ATLAS experimental measurements described in Sec.~\ref{subsec:exp_and_th}, we adopt $\eta_{\mathrm{gap}}=1$ and $p_T^{\mathrm{ref}} \in [0.5,2]$ GeV as default settings for all theoretical calculations. We trained GP emulators using 500 design points, which efficiently cover the model's parameter space~\cite{JETSCAPE:2020mzn}. These emulators allow us to generate statistically robust predictions and to propagate the uncertainty induced by the prior distribution of model parameters without requiring additional expensive event-by-event simulations. In this Section, we present the emulator's predictions as $90\%$ credible interval bands for three observables:
\begin{itemize}
    \item (i) the differential radial flow coefficient $v_0(p_T)$;
    \item (ii) its $p_T$-integrated counterpart $v_0$;
    \item (iii) the ratio $v_0(p_T)/v_0$, which isolates the shape information of the $p_T$ dependence independently of the overall normalization.    
\end{itemize}
We begin with the $p_T$-integrated observable $v_0$, which captures the overall magnitude of radial flow fluctuations. Fig.~\ref{fig:global_v0} shows global $v_0$ as a function of collision centrality for the Grad viscous correction scheme. The gray band (``\textsc{Jetscape} Prior 90\% C.I.'') represents the 90\% credible interval obtained by propagating the full prior distribution through the emulator. This band therefore reflects the theoretical uncertainty associated with the present range of model parameters before calibration. The relative standard deviation of the $p_T$-integrated $v_0$ increases toward more peripheral collisions. This trend indicates that peripheral events are more sensitive to variations in model parameters, particularly those controlling the initial state geometry and fluctuation strength. In central collisions, geometric fluctuations are partially averaged out due to the large number of participating nucleons, leading to a comparatively narrower prior band. In contrast, peripheral collisions amplify the impact of nucleon-level fluctuations and free-streaming dynamics, resulting in a broader spread of predictions. Superimposed on the prior band are predictions obtained from different calibration strategies described as follows: the blue band corresponds to a calibration including only $U(x_T)$~\cite{Domingues:2026mis} observables, while the orange band represents a calibration using only $p_T$-integrated observables within the \textsc{Jetscape} framework~\cite{JETSCAPE:2020mzn}. As expected, the posterior constrained by $p_T$-integrated observables (orange band) significantly reduces the uncertainty in global $v_0$ and provides an improved description of the ATLAS Pb+Pb $\sqrt{s_{NN}}=5.02$ TeV data. This behavior indicates the constraining power of integrated flow observables on the normalization of radial flow fluctuations. This plot shows the energy dependence of this observable, for more central regions, say 0--10\% we see that the prior range for 2.76 TeV cannot completely cover the experimental for 5.02 TeV as shown in the model-to-data bottom panel.    

\begin{figure}[H]
    \centering
    \includegraphics[width=\linewidth]{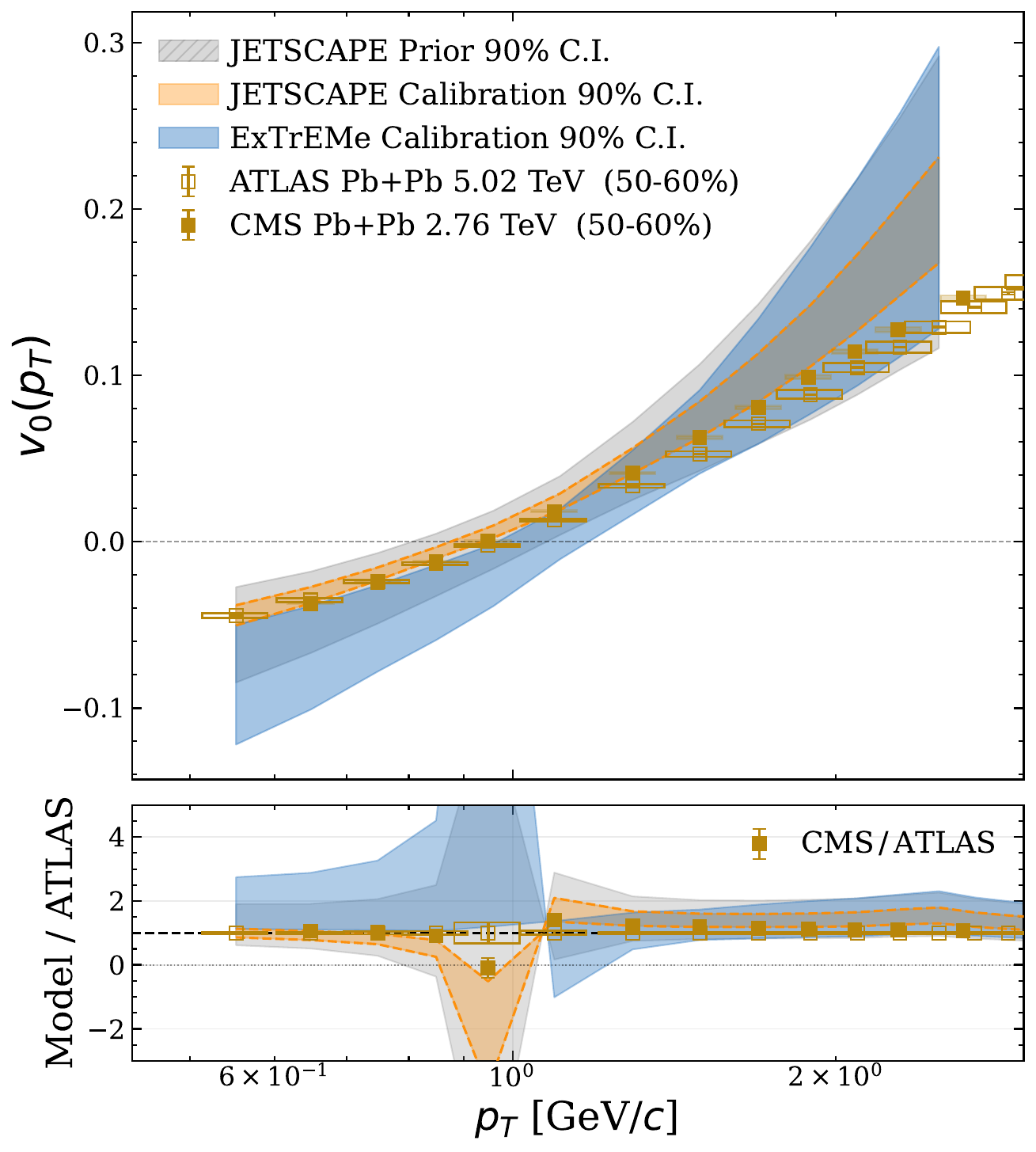}
    \caption{Emulator predictions for differential $v_0(p_T)$ in 50--60\% Pb+Pb collisions at $\sqrt{s_{NN}}=2.76~\mathrm{TeV}$. The gray band indicates the 90\% prior credible interval from~\cite{JETSCAPE:2020mzn}. The orange band corresponds to the posterior distribution calibrated on integrated observables~\cite{JETSCAPE:2020mzn}. The blue band represents the posterior calibrated on the scaled spectra~\cite{Domingues:2026mis}. Experimental ATLAS/CMS data are shown for comparison. The bottom panel highlights the model-to-data ratio.}
    \label{fig:v0pt_pred_60-70}
\end{figure}

In the following, we analyzed the differential observable $v_0(p_T)$, which provide complementary information on the momentum dependence of radial flow fluctuations.
Fig.~\ref{fig:v0pt_pred_60-70} shows emulator predictions in the 50--60\% centrality class for Pb--Pb collisions at $\sqrt{s_{NN}}=2.76$ TeV. The gray hatched band represents the 90\% credible interval obtained from the full prior parameter space. This band quantifies the intrinsic theoretical uncertainty before calibration and reflects the sensitivity of $v_0(p_T)$ to variations in model parameters. The substantial width of the prior interval, particularly at intermediate and high $p_T$, indicates that the differential observable retains significant sensitivity to model assumptions beyond those constrained by integrated quantities. Colored bands correspond to posterior predictions for different viscous correction schemes. In particular, the orange band represents the posterior distribution obtained in Ref.~\cite{JETSCAPE:2020mzn} using only $p_T$-integrated observables in the calibration. 
The corresponding posterior median is shown together with its credible interval. We observe that calibration to integrated observables significantly reduces the theoretical uncertainty relative to the prior band, demonstrating that global constraints propagate nontrivially to the differential observable. However, residual deviations between model and data at higher $p_T$ suggest that the momentum dependence of radial flow fluctuations provides additional constraining power beyond integrated measurements. The lower panel displays the ratio of model predictions to experimental data, allowing for a more quantitative assessment of agreement across the $p_T$ range. Within uncertainties, the posterior predictions provide a reasonable description of the data in the low- and intermediate-$p_T$ region, while tensions become more visible at larger transverse momentum, where nonflow effects or limitations of the viscous correction scheme may play a role.
Prior model predictions were obtained by sampling the full parameter space uniformly within the predefined design ranges used for emulator training. For each viscous correction scheme, we generate $2,000$ independent parameter points,
\begin{equation}
\boldsymbol{\theta}_i \sim \mathcal{U}(\boldsymbol{\theta}_{\min}, \boldsymbol{\theta}_{\max}),
\label{eq:prior_uniform_sampling}
\end{equation}
where $\boldsymbol{\theta}_{\min}$ and $\boldsymbol{\theta}_{\max}$ denote the lower and upper bounds of the prior ranges, respectively. We refer readers to Table I in Ref.~\cite{Domingues:2026mis} for details about the prior intervals and model parameters. This sampling strategy ensures a consistent and unbiased exploration of the physically admissible parameter space.

For each sampled parameter vector, the trained emulators are evaluated to predict the principal component coefficients associated with the PCA representation of $v_{0}$, $v_{0}(p_{T})$, and $v_{0}(p_{T})/v_0$. Emulator uncertainties are retained at the level of the principal components and propagated to observable space, as described below.

\begin{figure*}[ht!]
    \centering
    
    \begin{subfigure}[b]{0.48\textwidth}
        \centering
        \includegraphics[width=\linewidth]{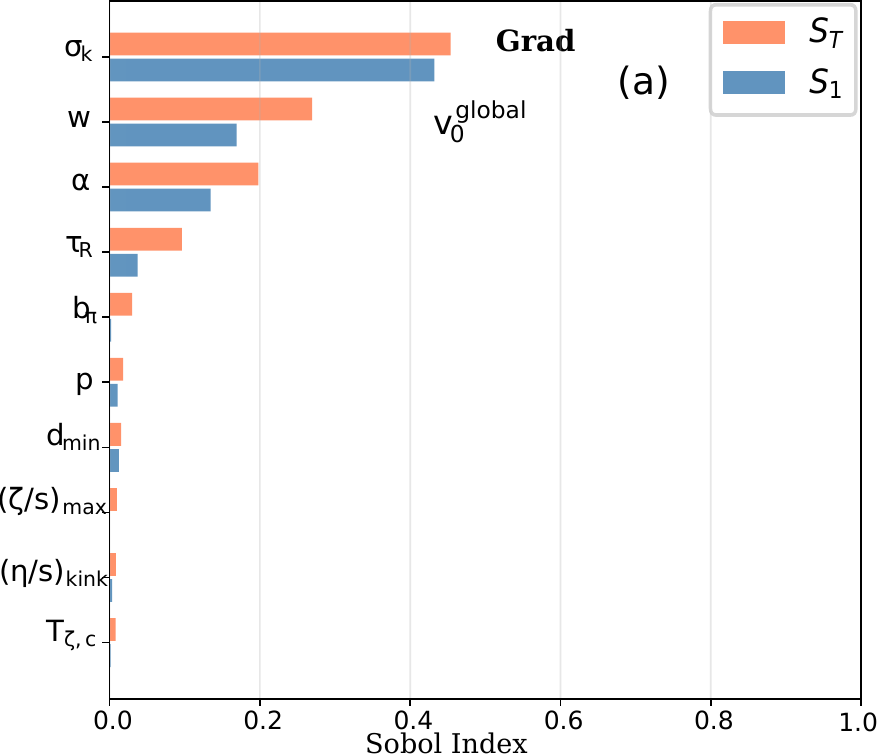}
        \label{fig:gsa_v0_Grad}
    \end{subfigure}
    \begin{subfigure}[b]{0.48\textwidth}
        \centering
        \includegraphics[width=\linewidth]{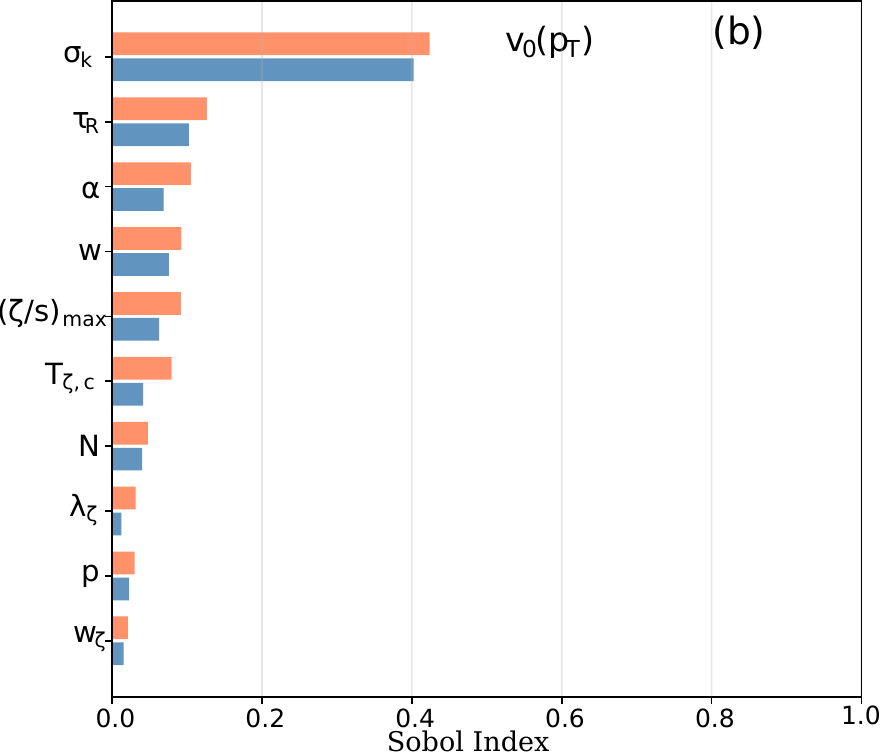}
        \label{fig:gsa_v0_ratio_Grad}
    \end{subfigure}
    \caption{
    Global sensitivity analysis under the Grad viscous correction prescription.
    Panel~(a) shows the ten most influential parameters for the integrated
    observable $v_0$, ranked by their total-order Sobol index $S_T$
    (orange) and first-order index $S_1$ (blue), computed independently
    for each retained PC coefficient and shown here for the leading PC.
    Panel~(b) shows the analogous ranking for the differential observable
    $v_0(p_T)$. In both panels, initial-state parameters dominate the
    sensitivity spectrum: the nucleon fluctuation parameter $\sigma_k$ and
    the nucleon width $w$ emerge as the leading sources of variance, followed by the free-streaming parameters $\tau_R$ and $\alpha$.
    The bulk viscosity peak $(\zeta/s)_{\rm max}$ acquires comparatively
    larger relative importance for $v_0(p_T)$ than for $v_0$,
    reflecting the sensitivity of the spectral shape to bulk viscous
    corrections at particlization.
    }
    \label{fig:gsa_Grad}
\end{figure*}

The results presented in this work focus on a representative mid-peripheral centrality class, $50$--$60\%$ and $60$--$70\%$, where radial flow effects remain significant while nonflow contributions are reduced. Predictions are shown as functions of transverse momentum, using a logarithmic scale in $p_T$ to emphasize the low-momentum region where radial flow effects are most pronounced. The model is only emulated up to $p_T = 2.5$~GeV/$c$, theoretical predictions are restricted to this range, even though the experimental data extend to higher $p_T$. We train emulators for all four viscous correction prescriptions; however, we show only the Grad results in this paper, since the differences among the prescriptions are negligible for the observables studied here. The corresponding figures for the other prescriptions are available in the public repository~\cite{data_github_page}. Furthermore, to understand which parameters are the most important for radial flow expansion, we performed a global sensitivity analysis, calculating all Sobol indices as depicted in Fig.~\ref{fig:gsa_Grad}.
We perform variance-based global sensitivity analysis using Sobol indices to identify the most influential model parameters for the integrated observable $v_0$, the differential coefficient $v_0(p_T)$, and their ratio $v_0(p_T)/v_0$. To our knowledge, this is the 
first systematic parameter-space sensitivity analysis carried out for these radial flow  observables over the full high-dimensional prior of a state-of-the-art hybrid heavy-ion model; the only precedent is the sensitivity analysis of $\delta p_T/\langle p_T\rangle$  reported in the \textsc{Jetscape} multisystem analysis~\cite{JETSCAPE:2020mzn}, to which $v_0$ is directly related.

The analysis reveals that the nucleon fluctuation parameter $\sigma_k$ is the leading source of variance for $v_0$, followed by the nucleon width $w$, the free-streaming exponent $\alpha$, and the free-streaming time scale $\tau_R$. We remind readers that the parameter $\sigma_k$ controls the width of the $\Gamma$-distribution used to modulate each participant nucleon's contribution to the 
initial energy density~\cite{JETSCAPE:2020mzn}; larger event-by-event fluctuations of the nucleon weights directly translate into larger fluctuations of the mean transverse momentum, which is what $v_0 \equiv \sigma[p_T]/\langle p_T\rangle$ measures. The nucleon width $w$ sets the spatial extent of the nucleon thickness function (Eq.~(6) of Ref.~\cite{JETSCAPE:2020mzn}): a narrower nucleon profile generates larger local energy-density gradients and thus stronger event-by-event fluctuations of the radial push. The free-streaming parameters $\tau_R$ and $\alpha$ govern the duration and energy-density dependence of the pre-equilibrium stage (Eq.~(19) of Ref.~\cite{JETSCAPE:2020mzn}); longer free-streaming times allow the system to build larger transverse flow before hydrodynamics begins, increasing $v_0$. Notably, $v_0$ exhibits relatively low sensitivity to the bulk viscosity parameters, consistent with its role as an integrated measure of radial flow fluctuations rather than a probe of the momentum shape, as shown in Fig.~\ref{fig:gsa_Grad} panel (a). 

For the differential observable $v_0(p_T)$ shown in Fig.~\ref{fig:gsa_Grad} panel (b), the dominant parameters shift to $\sigma_k$, $\tau_R$, $\alpha$, $w$, and the bulk viscosity peak $(\zeta/s)_{\rm max}$. The appearance of $(\zeta/s)_{\rm max}$ among the top contributors reflects the well-known sensitivity of the spectral shape to bulk viscous effects: bulk viscosity acts to isotropize the local momentum distribution at particlization, suppressing high-$p_T$ yields and redistributing particles toward lower momenta, thereby modifying the covariance structure of $v_0(p_T)$ across the $p_T$ range~\cite{Schenke:2020uqq, Du:2025dpu, Jia:2025rab}. The enhanced role of $\tau_R$ relative to $v_0$ is consistent with the expectation that the momentum dependence of radial flow fluctuations retains memory of the early-time dynamics: the duration of free-streaming affects not only the magnitude but also the shape of the radial push imparted to the system before hydrodynamics takes over.
\subsection{Scaling of the radial flow $v_0(p_T)$}
\label{subsec:scaled_v0pt}
\begin{figure}
    \centering
    \includegraphics[width=\linewidth]{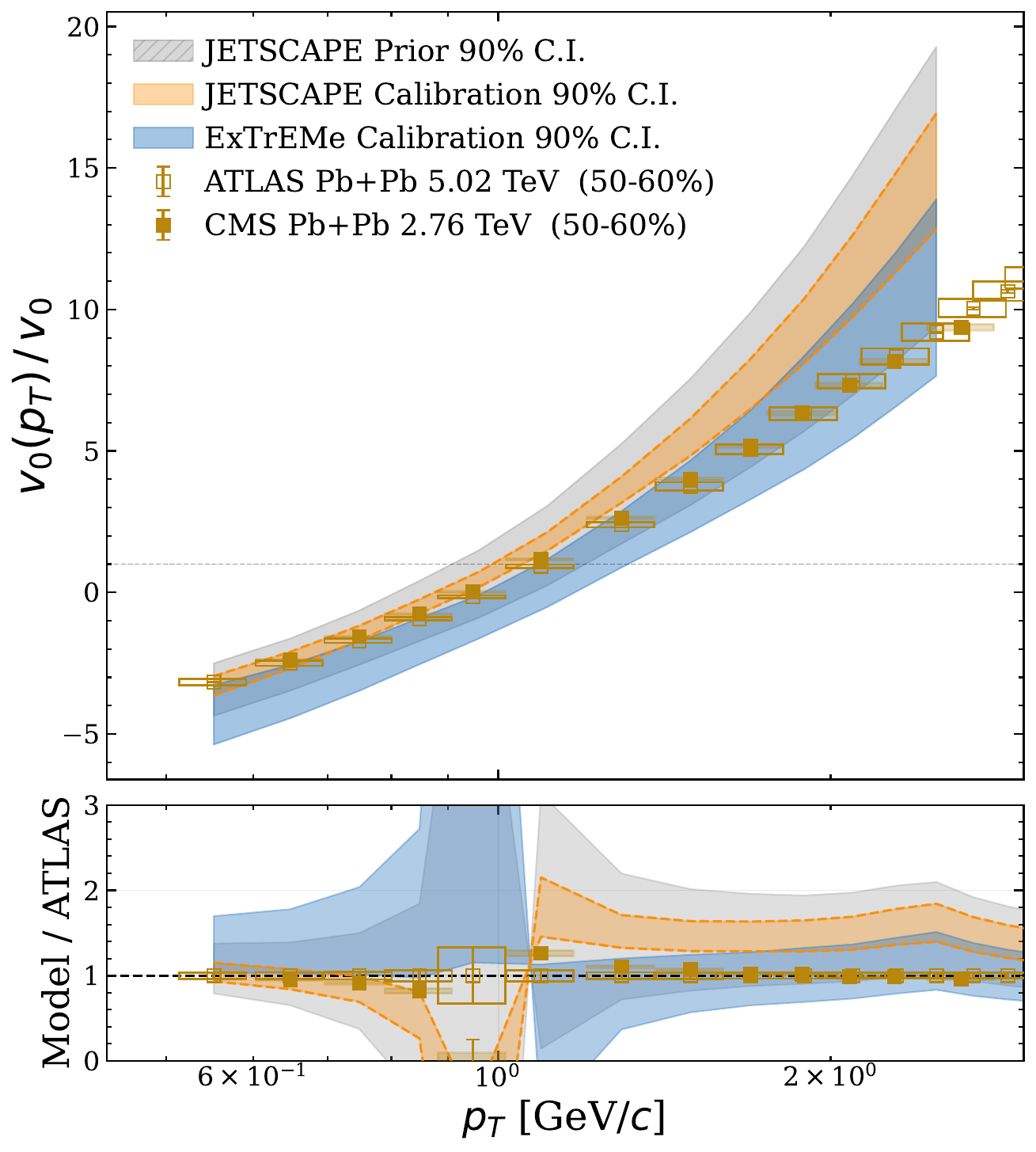}
    \caption{Emulator predictions for the scaled radial flow $v_0(p_T)/v_0$ in 50--60\% Pb+Pb collisions at $\sqrt{s_{NN}} = 2.76$ TeV. The gray band represents the 90\% credible interval for the full prior parameter space~\cite{JETSCAPE:2020mzn}. The orange band corresponds to the posterior distribution calibrated on integrated observables~\cite{JETSCAPE:2020mzn}. The blue band represents the posterior calibrated on the scaled spectra~\cite{Domingues:2026mis}. The lower panel shows the ratio of the model prediction to the CMS and ATLAS data.}
    \label{fig:v0_ratio_Grad}
\end{figure}

To investigate the momentum dependence of radial flow and minimize common systematic uncertainties, we analyze the scaled observable $v_0(p_T)/v_0$. This ratio provides a robust constraint on the transverse momentum profile of the expansion, independent of the overall flow magnitude. Fig.~\ref{fig:v0_ratio_Grad} displays the comparison between the Grad viscous correction scheme for $\sqrt{s_{NN}} = 2.76$ TeV and ATLAS experimental data for Pb+Pb collisions at $\sqrt{s_{NN}} = 5.02$ TeV within the 50--60\% centrality interval. The emulator predictions, represented by the colored bands, 90\% credible intervals indicates good agreement with the measured data points across the $p_T$ range. Although the model calibration was performed primarily using 2.76 TeV data~\cite{JETSCAPE:2020mzn, Domingues:2026mis}, the framework successfully describes measurements at 5.02 TeV, confirming a degree of universality in the underlying hydrodynamic description across different collision energies. This cross-energy consistency suggests that the Grad viscous correction can captures the essential physics of radial flow evolution.
The success in reproducing the ratio $v_0(p_T)/v_0$ at a different collision energy than that used for calibration suggests an approximate universality across collision energies, as already pointed out by previous works~\cite{Schenke:2020uqq, Parida:2024ckk, ATLAS:2025ztg, Du:2025dpu}.
\begin{figure}
    \centering
    \includegraphics[width=\linewidth]{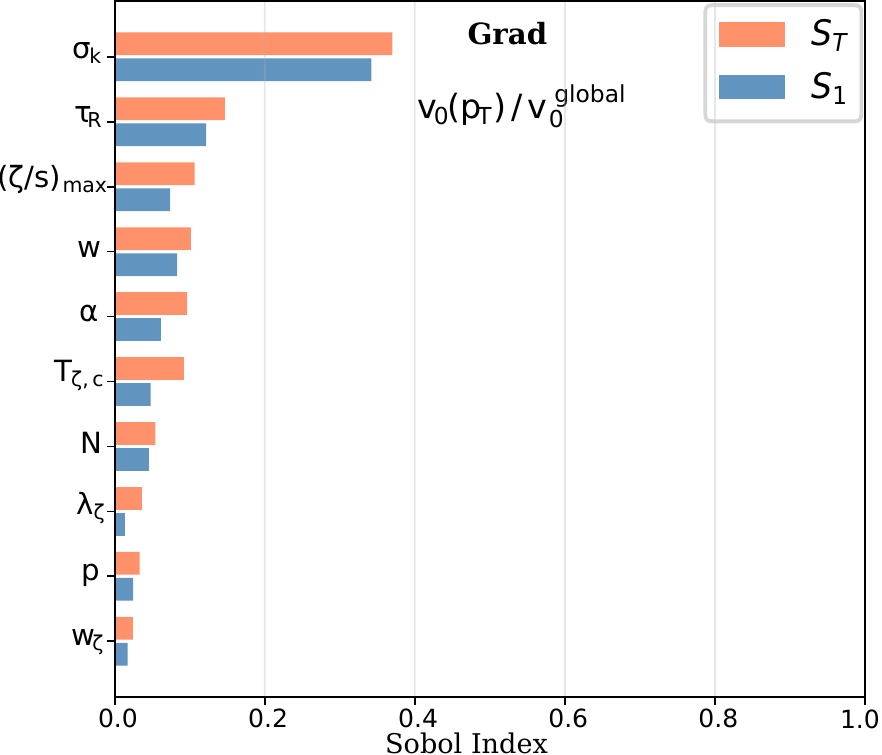}
    \caption{Global sensitivity analysis for the scaled ratio $v_0(p_T)/v_0$ under the Grad viscous correction prescription. Bars show the ten most influential parameters ranked by their total-order Sobol index $S_T$ (orange) and first-order index $S_1$ (blue). The nucleon fluctuation parameter $\sigma_k$ and the free-streaming time scale $\tau_R$ remain the leading drivers, while the bulk viscosity peak $(\zeta/s)_{\rm max}$ acquires comparatively larger relative importance than in $v_0$ and $v_0(p_T)$, consistent with earlier studies identifying $v_0(p_T)/v_0$ as a preferred probe of bulk viscous corrections~\cite{Schenke:2020uqq, Parida:2024ckk, ATLAS:2025ztg, Du:2025dpu}.
    }
    \label{fig:v_0_ratio_gsa}
\end{figure}
For the scaled ratio shown in Fig.~\ref{fig:v_0_ratio_gsa}, the sensitivity ordering becomes $\sigma_k$, $\tau_R$, $(\zeta/s)_{\rm max}$, $w$, and $\alpha$. The most notable feature is the \emph{increased} relative importance of $(\zeta/s)_{\rm max}$ compared to $v_0$ and $v_0(p_T)$. Because the ratio $v_0(p_T)/v_0$ isolates the shape of the momentum dependence while removing the overall normalization, common initial-state fluctuations encoded in $\sigma_k$ and $w$ partially cancel in the ratio, leaving bulk viscous corrections as a comparatively more prominent driver. This result is consistent with earlier hydrodynamic studies that identified $v_0(p_T)/v_0$ as particularly sensitive to bulk viscosity~\cite{Schenke:2020uqq, Parida:2024ckk, ATLAS:2025ztg, Du:2025dpu}, reinforcing the view that the scaled ratio constitutes a cleaner probe of transport coefficients than the absolute observable. The free-streaming parameters $\tau_R$ and $\alpha$ retain significant influence, indicating that the shape of the radial flow profile is set during the pre-equilibrium stage and is not fully erased by the subsequent hydrodynamic evolution.
\section{CONCLUSIONS}
\label{sec:conclusions}
We have presented a systematic investigation of the differential radial flow coefficient $v_0(p_T)$, its $p_T$-integrated counterpart $v_0$, and the scaled ratio $v_0(p_T)/v_0$ in Pb+Pb collisions at $\sqrt{s_{NN}} = 2.76~\mathrm{TeV}$, using CMS open data for mid-peripheral centrality classes of 50--60\% and 60--70\%.

On the experimental side, the measured $v_0(p_T)$ distributions exhibit a clear energy dependence in their overall magnitude when compared with ATLAS measurements at $\sqrt{s_{NN}} = 5.02~\mathrm{TeV}$. However, upon scaling by the integrated observable $v_0$, this energy dependence is largely removed, and the resulting ratio $v_0(p_T)/v_0$ displays an approximate universality that is nearly independent of both collision energy and centrality at low $p_T$, persisting up to $p_T\sim3~\mathrm{GeV}$. This behavior suggests that the integrated $v_0$ captures most of the energy-dependent information, while the momentum dependence of radial flow fluctuations is governed by an underlying mechanism that is largely collision-energy independent within the explored kinematic range. The factorization assumption relating $v_0(p_T)$ to the normalized transverse-momentum covariance and the integrated $v_0$ was validated across different $p_T^{\mathrm{ref}}$ ranges and pseudorapidity gap values, confirming its robustness in the low-$p_T$ regime.

On the theoretical side, we trained emulators for $v_0$ and $v_0(p_T)$ and used them to generate predictions with parameter posteriors from previous Bayesian calibrations of the model, which did not use $v_0$, $v_0(p_T)$, or $v_0(p_T)/v_0$ in the calibration. These are the posteriors calibrated on $p_T$-integrated observables~\cite{JETSCAPE:2020mzn} and the posteriors calibrated on the scaled spectra~\cite{Domingues:2026mis}. They provide a reasonable description of the experimental data in the low- and intermediate-$p_T$ regime. Notably, parameters calibrated at 2.76~TeV also describe the ATLAS data at 5.02~TeV for the ratio $v_0(p_T)/v_0$, providing further evidence for the near universality of the radial flow momentum dependence across collision energies.

A global variance-based sensitivity analysis using Sobol indices was performed to identify the model parameters that most strongly govern the differential and integrated radial flow dynamics. The nucleon fluctuation parameter $\sigma_k$ emerges as the dominant parameter across all viscous correction schemes considered, while the nucleon width $w$, the free-streaming exponent $\alpha$, and the free-streaming time scale $\tau_R$ also play significant roles. Importantly, the integrated observable $v_0$ exhibits relatively low sensitivity to bulk viscosity parameters, in contrast to the differential observable $v_0(p_T)$, highlighting the complementary nature of these two probes in constraining the transport properties of the QGP. A natural extension of this analysis, which we defer to future work, is to rescale the $p_T$ axis by the mean transverse momentum, studying $v_0(p_T) /v_0$ against $p_T / \langle p_T \rangle$. Such a rescaling is expected to further suppress the sensitivity to initial-state parameters and amplify the relative contribution of bulk viscous corrections, potentially making this observable an even more direct probe of the bulk viscosity of the QGP~\cite{Schenke:2020uqq, Parida:2024ckk, ATLAS:2025ztg, Du:2025dpu}.

Future extensions of this work include the application of $v_0(p_T)$ as a direct calibration observable in a full Bayesian inference framework, the extension to more central collision classes and additional collision systems, and the investigation of the interplay between radial and anisotropic flow within an unified emulation strategy. The present results indicates that $v_0(p_T)$ constitutes a powerful and complementary probe of the collective dynamics and transport properties of the hot QCD medium formed in ultrarelativistic heavy-ion collisions.
\section*{Acknowledgments}
We thank the \textsc{Jetscape} Collaboration for access to their hybrid simulation framework and analysis scripts, and the CMS Collaboration for making their data publicly available through the CERN Open Data Portal. A. E. F. G. F acknowledges the Brazilian Federal Agency for Support and Evaluation of Graduate Education (CAPES) for the support through a MSc fellowship (File No. 88887.372908/2026-00). T.S.D. acknowledges financial support from the Brazilian National Council for Scientific and Technological Development (CNPq) through a PhD fellowship under process No.\ 141432/2025-0, INCT-FNA grant number 464898/2014-5, and from the CAPES scholarship under process No. 88881.220538/2025-01. M.L. was supported by FAPESP projects 2017/05685-2 and 2018/24720-6, and by project INCT-FNA Proc. No. 464898/2014-5. C. A. B. acknowledges funding from CNPq (Proc. No. 309962/2023-4) and FAPESP (Proc. No. 2018/25225-9).
\textit{Use of AI tools.} The authors used the large language models Claude Fable~5, Claude Opus~5.1, and Claude Opus~5.5 (Anthropic) to help improve the clarity of the manuscript text and to assist in writing the plotting scripts. The authors reviewed and edited all AI-assisted material and take full responsibility for the content of this paper, including the scientific results and conclusions.
\section*{Data Availability}
This study uses both theoretical simulation data generated by the authors and publicly available experimental data. The CMS data are available from the CERN Open Data Portal~\cite{cms_open_data}, and the ATLAS measurements are available from HEPData~\cite{ATLAS:2025ztg}. The simulated datasets, trained emulators, design points, parameter ranges, and analysis scripts needed to reproduce the results are available in the GitHub repository associated with this work~\cite{data_github_page}.
\appendix
\label{sec:appendix}
\section{Principal Component Analysis}
\label{appendix:appendix_dpts_and_pca}
To efficiently reduce the dimensionality of the analyzed observables while preserving their dominant physical features, we perform a PCA on the model calculations obtained across all design points and viscous correction prescriptions. This procedure enables the construction of accurate surrogate models using a reduced latent representation of the observables while maintaining high reconstruction fidelity.

\begin{figure}[H]
    \centering
    \includegraphics[width=\linewidth]{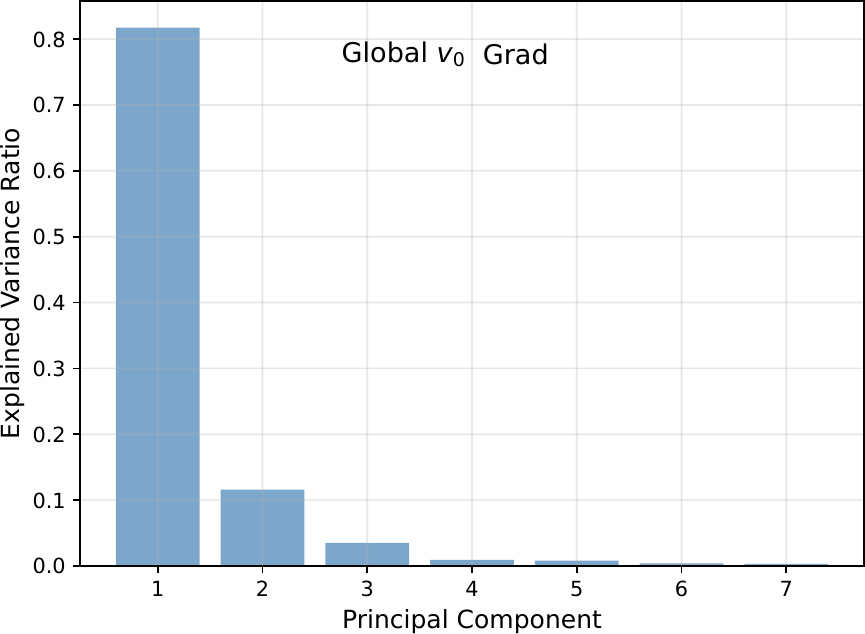}
    \caption{Explained variance ratio of the principal component decomposition for global $v_0$ using the Grad viscous correction. With only 3 PCs we can explain more than $95\%$ cumulative variance over all the $N_{\mathrm{bins}}$.}
    \label{fig:pca_Grad}
\end{figure}

For each $\delta f$ model (Grad, CE, PTM, and PTB), we build a data matrix from the model calculations of the observables
\begin{equation}
\mathcal{O}_m \in \left\{
v_0,\,
v_0(p_T)
\right\}.
\label{eq:observables}
\end{equation}
For a given design point indexed by $i$, the observable vector is defined as
\begin{equation}
\mathbf{y}^{(m)}_i
=
\left\{
\mathcal{O}_m^{(c_k)}(p_{T,j})
\right\},
\qquad
\begin{aligned}
j &= 1,\ldots,N_p,\\
k &= 1,\ldots,N_c,
\end{aligned}
\label{eq:pca_analysis}
\end{equation}
where $N_p$ denotes the number of transverse momentum bins and $N_c$ the number of centrality classes. The superscript $(c_k)$ labels the centrality interval, while $p_{T,j}$ corresponds to the center of the $j$-th transverse momentum bin. The observable vector $\mathbf{y}^{(m)}_i$ therefore contains the full set of values for observable $\mathcal{O}_m$ across all momentum and centrality bins for the $i$-th design point. The full design matrix is then assembled as
\begin{equation}
\mathbf{Y}^{(m)}
=
\begin{pmatrix}
\mathbf{y}^{(m)}_1 \\
\mathbf{y}^{(m)}_2 \\
\vdots \\
\mathbf{y}^{(m)}_{N_{\mathrm{design}}}
\end{pmatrix},
\label{eq:design_matrix}
\end{equation}
with dimensions $N_{\mathrm{design}} \times N_{\mathrm{bins}}$, where $N_{\mathrm{bins}} = N_p \times N_c$.  
Prior to performing PCA, each observable component is standardized to a zero mean and unit variance according to
\begin{equation}
\widetilde{Y}^{(m)}_{ij}
=
\frac{
Y^{(m)}_{ij}
-
\mu^{(m)}_j
}{
\sigma^{(m)}_j
},
\label{eq:standardization}
\end{equation}
where $\mu^{(m)}_j$ and $\sigma^{(m)}_j$ are respectively the mean and standard deviation of the $j$-th observable component evaluated over all design points. This preprocessing step ensures that all momentum and centrality bins contribute comparably to the decomposition and prevents bins with larger absolute magnitudes from dominating the principal components. The standardized matrix $\widetilde{\mathbf{Y}}^{(m)}$ is decomposed through a singular value decomposition (SVD)~\cite{doi:10.1137/1.9781421407944},
\begin{equation}
\widetilde{\mathbf{Y}}^{(m)}
=
\mathbf{U}^{(m)}
\mathbf{S}^{(m)}
\left(
\mathbf{V}^{(m)}
\right)^{\mathrm T},
\label{eq:singular_value_decomposition}
\end{equation}
where $\mathbf{U}$ contains the left singular vectors, $\mathbf{S}$ is a diagonal matrix of singular values $s_k$, and $\mathbf{V}$ contains the right singular vectors corresponding to the orthonormal directions in observable space. The SVD decomposition was computed using the Numpy ecosystem~\cite{harris2020numpy}. The explained variance ratio associated with the $k$-th principal component is given by
\begin{equation}
\lambda_k^{(m)}
=
\frac{
\left(s_k^{(m)}\right)^2
}{
\sum_{\ell}
\left(s_\ell^{(m)}\right)^2
}.
\label{eq:explained_variance_ratio}
\end{equation}
Rather than fixing the number of retained principal components \textit{a priori}, we determine the minimum number $N_{\mathrm{PC}}$ required such that the cumulative explained variance satisfies
\begin{equation}
\sum_{k=1}^{N_{\mathrm{PC}}}
\lambda_k^{(m)}
\geq
0.90,
\label{eq:PCA_threshold}
\end{equation}
thereby preserving at least $90\%$ of the total variance of the original observable space. The principal component scores associated with the $i$-th design point are defined as
\begin{equation}
\mathbf{z}^{(m)}_i
=
\sqrt{N_{\mathrm{design}}-1}\,
\mathbf{U}^{(m)}_{i,\,1:N_{\mathrm{PC}}},
\label{eq:pca_scores2}
\end{equation}
which constitute the reduced latent target variables. For the global observable $v_0$, the PCA analysis shows that the first five PCs capture more than $99\%$ of the total variance, as illustrated in Fig.~\ref{fig:pca_Grad}. In contrast, the differential observable $v_0(p_T)$ exhibits broader variance across centrality bins and requires at least nine principal components to achieve the $90\%$ cumulative threshold over all the viscous corrections. For more details, we refer the reader to the GitHub repository associated with this manuscript~\cite{data_github_page}.
To quantify the truncation error introduced by retaining $N_{\mathrm{PC}}$ components, we reconstruct the physical observable matrix $\widehat{\mathbf{Y}}^{(m)}$ from the truncated scores via
\begin{equation}
\widehat{Y}^{(m)}_{ij} = \mu^{(m)}_j + \sigma^{(m)}_j \sum_{k=1}^{N_{\mathrm{PC}}} Z^{(m)}_{ik} V^{(m)}_{jk},
\end{equation}
where $Z^{(m)}_{ik} = \sqrt{N_{\mathrm{design}}-1}^{-1} U^{(m)}_{ik} S^{(m)}_{kk}$. We evaluate the root-mean-square relative reconstruction error across all design points and kinematic bins:
\begin{equation}
\epsilon_{\mathrm{recon}}^{(m)} = \sqrt{\frac{1}{N_{\mathrm{design}} N_{\mathrm{bins}}} \sum_{i=1}^{N_{\mathrm{design}}} \sum_{j=1}^{N_{\mathrm{bins}}} \left( \frac{\widehat{Y}^{(m)}_{ij} - Y^{(m)}_{ij}}{Y^{(m)}_{ij}} \right)^2}.
\label{eq:recon_error}
\end{equation}
Across all four viscous correction schemes, $\epsilon_{\mathrm{recon}}$ remains below $0.7\%$ for $v_0$ and below $1.1\%$ for $v_0(p_T)$. This negligible reconstruction error confirms that $N_{\mathrm{PC}}$ principal components capture the dominant physical variations without losing relevant physical information.
\section{Emulator Validation}
\label{appendix:emu_validation}
\begin{figure}
    \centering
    \includegraphics[width=\linewidth]{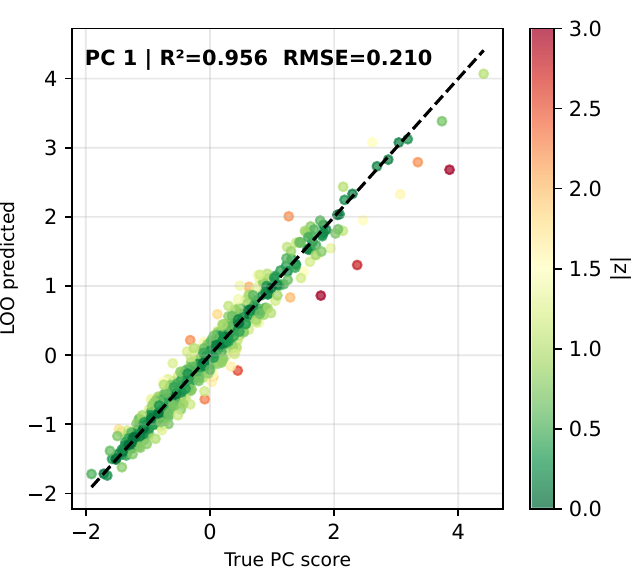}
    \caption{LOO cross-validation scatter plots for the GP emulators of the integrated $v_0$ observable using the Grad viscous correction prescription. Here, we show the true PC score versus the LOO-predicted score for the first principal component. Points are color-coded by the standardized residual $|y - \hat{y}|/\sigma$; a perfect emulator lies on the dashed diagonal. The $R^2$ and RMSE values, together with the empirical $1\sigma$ and $2\sigma$ coverage fractions, are reported.}
    \label{fig:emu_validation_Grad_v_0}
\end{figure}

To assess the predictive accuracy of the GP emulators, we perform a Leave-One-Out (LOO) cross-validation directly in the latent PC-score space~\cite{Domingues:2026mis}. For each design point $i$, we retrain the emulator on all remaining points and predict the held-out PC scores $\widehat{z}_{ik}$. For a GP trained on inputs $\mathbf{X}$ with kernel matrix $\mathbf{K}$ and noise variance $\sigma_n^2$, the LOO prediction and its uncertainty for the $i$-th point are
\begin{equation}
    \widehat{z}_{ik}^{\,\mathrm{LOO}} = z_{ik} - \frac{\left[\mathbf{K}_n^{-1}\,\mathbf{z}_k\right]_i}{\left[\mathbf{K}_n^{-1}\right]_{ii}},
    \qquad
    \sigma_{ik}^{\,\mathrm{LOO}} = \frac{1}{\sqrt{\left[\mathbf{K}_n^{-1}\right]_{ii}}},
    \label{eq:loo_analytical}
\end{equation}
where $\mathbf{K}_n = \mathbf{K} + \sigma_n^2\,\mathbf{I}$ is the noise-regularized kernel matrix. The quality of the LOO predictions is quantified for each principal component $k$ through the coefficient of determination
\begin{equation}
    R^2_k = 1 - \frac{\displaystyle\sum_i \bigl(z_{ik} - \widehat{z}_{ik}^{\,\mathrm{LOO}}\bigr)^2}{\displaystyle\sum_i \bigl(z_{ik} - \langle z_k\rangle\bigr)^2},
    \label{eq:r2_score}
\end{equation}
together with the root-mean-square error (RMSE) and coverage fractions at the $1\sigma$ and $2\sigma$ levels. A well-calibrated emulator should yield $68\%$ and $95\%$ coverage fractions matching the nominal Gaussian values, while $R^2_k \gtrsim 0.95$ indicates that the GP captures nearly all variance in the PC scores. Fig.~\ref{fig:emu_validation_Grad_v_0} shows the LOO scatter plots for the integrated $v_0$ observable under the Grad viscous correction prescription for the first PC displaying the true score on the horizontal axis versus the LOO-predicted score on the vertical axis, with points color-coded by the standardized residual $|z_{ik} - \widehat{z}_{ik}^{\,\mathrm{LOO}}|/\sigma_{ik}^{\,\mathrm{LOO}}$. Analogous figures for the CE, PTM, and PTB viscous corrections are collected in the public repository associated with this work~\cite{data_github_page} and are not shown here. The same validation procedure was implemented for emulators trained in $v_0(p_T)$. 

Across all four viscous correction prescriptions, the first PCs achieve $R^2_k > 0.95$ on the LOO metric, with coverage fractions within a few percent of the nominal $68\%$ and $95\%$ values. The pull distributions are approximately centered and have standard deviations close to unity, indicating that the GP uncertainty estimates are well calibrated. This confirm that the emulators provide a high-fidelity and reliable replacement for the full hybrid model throughout the parameter space.
\bibliography{ref.bib}

@book{bootstrap,
  title={Asymptotic Theory for Bootstrap Methods in Statistics},
  author={Beran, R.J. and Ducharme, G.R. and Universit{\'e} de Montr{\'e}al. Centre de recherches math{\'e}matiques},
  isbn={9782921120104},
  series={Les Publications CRM},
  url={https://books.google.com.br/books?id=XQvvAAAAMAAJ},
  year={1991},
  publisher={Les Publications CRM}
}

@article{ATLAS:2025ztg,
    author = "Aad, Georges and others",
    collaboration = "ATLAS",
    title = "{Evidence for the Collective Nature of Radial Flow in Pb+Pb Collisions with the ATLAS Detector}",
    eprint = "2503.24125",
    archivePrefix = "arXiv",
    primaryClass = "nucl-ex",
    reportNumber = "CERN-EP-2025-068",
    doi = "10.1103/ldcn-r2lq",
    journal = "Phys. Rev. Lett.",
    volume = "136",
    number = "3",
    pages = "032301",
    year = "2026"
}

@article{ALICE:2025iud,
    author = "Acharya, Shreyasi and others",
    collaboration = "ALICE",
    title = "{Long-range transverse momentum correlations and radial flow in Pb$-$Pb collisions at the LHC}",
    eprint = "2504.04796",
    archivePrefix = "arXiv",
    primaryClass = "nucl-ex",
    reportNumber = "CERN-EP-2025-084",
    doi = "10.1103/l36g-6f46",
    journal = "Phys. Rev. Lett.",
    volume = "136",
    number = "3",
    pages = "032302",
    year = "2026"
}

@misc{atlas-qm25,
    author = {Somadutta Bhatta},
    title  = {Evidence For the Collective Nature of Radial Flow in Pb+Pb Collisions with ATLAS},
    year   = {2025},
    note   = {Quark Matter 2025. \url{https://indico.cern.ch/event/1334113/contributions/6394983/}}
}

@misc{atlas-is25,
    author={Somadutta Bhatta},
    title  = {Recent Flow and Correlation Measurements in Large and Small Collision Systems with ATLAS},
    year   = {2025},
    note   = {Initial Stages 2025. \url{https://indico.cern.ch/event/1479384/contributions/6663079/}}
}

@misc{cms_open_data,
  author       = {{CMS collaboration}},
  title        = {{HIMinBiasUPC primary dataset in RECO format from the 2.76 TeV Pb-Pb run of 2011 (/HIMinBiasUPC/HIRun2011-12Jun2013-v1/RECO)}},
  year         = {2020},
  howpublished = {CERN Open Data Portal},
  doi          = {10.7483/OPENDATA.U18S.LSSA},
  url          = {https://doi.org/10.7483/OPENDATA.U18S.LSSA}
}

@misc{data_github_page,
  howpublished = {\url{https://github.com/ThiagoSDomingues/JETSCAPE_studies/tree/main/CMS_v0_analysis}}
}

@article{Schenke:2020uqq,
    author = {Schenke, Bj{\"o}rn and Shen, Chun and Teaney, Derek},
    title = "{Transverse momentum fluctuations and their correlation with elliptic flow in nuclear collision}",
    eprint = "2004.00690",
    archivePrefix = "arXiv",
    primaryClass = "nucl-th",
    doi = "10.1103/PhysRevC.102.034905",
    journal = "Phys. Rev. C",
    volume = "102",
    number = "3",
    pages = "034905",
    year = "2020"
}

@article{Parida:2024ckk,
    author = "Parida, Tribhuban and Samanta, Rupam and Ollitrault, Jean-Yves",
    title = "{Probing collectivity in heavy-ion collisions with fluctuations of the pT spectrum}",
    eprint = "2407.17313",
    archivePrefix = "arXiv",
    primaryClass = "nucl-th",
    doi = "10.1016/j.physletb.2024.138985",
    journal = "Phys. Lett. B",
    volume = "857",
    pages = "138985",
    year = "2024"
}

@article{Du:2025dpu,
    author = "Du, Lipei",
    title = "{Characterizing radial flow fluctuations in relativistic heavy-ion collisions at top RHIC and LHC energies}",
    eprint = "2508.07184",
    archivePrefix = "arXiv",
    primaryClass = "nucl-ex",
    doi = "10.1103/9pvq-ph1c",
    journal = "Phys. Rev. C",
    volume = "113",
    number = "1",
    pages = "014901",
    year = "2026"
}

@article{Gardim:2019iah,
    author = "Gardim, Fernando G. and Grassi, Fr{\'e}d{\'e}rique and Ishida, Pedro and Luzum, Matthew and Ollitrault, Jean-Yves",
    title = "{$p_T$-dependent particle number fluctuations from principal-component analyses in hydrodynamic simulations of heavy-ion collisions}",
    eprint = "1906.03045",
    archivePrefix = "arXiv",
    primaryClass = "nucl-th",
    doi = "10.1103/PhysRevC.100.054905",
    journal = "Phys. Rev. C",
    volume = "100",
    number = "5",
    pages = "054905",
    year = "2019"
}

@article{Roch:2024xhh,
    author = "Roch, Hendrik and Jahan, Syed Afrid and Shen, Chun",
    title = "{Model emulation and closure tests for (3+1)D relativistic heavy-ion collisions}",
    eprint = "2405.12019",
    archivePrefix = "arXiv",
    primaryClass = "nucl-th",
    doi = "10.1103/PhysRevC.110.044904",
    journal = "Phys. Rev. C",
    volume = "110",
    number = "4",
    pages = "044904",
    year = "2024"
}

@article{Bernhard:2019bmu,
    author = "Bernhard, Jonah E. and Moreland, J. Scott and Bass, Steffen A.",
    title = "{Bayesian estimation of the specific shear and bulk viscosity of quark{\textendash}gluon plasma}",
    doi = "10.1038/s41567-019-0611-8",
    journal = "Nature Phys.",
    volume = "15",
    number = "11",
    pages = "1113--1117",
    year = "2019"
}

@article{JETSCAPE:2020mzn,
    author = "Everett, D. and others",
    collaboration = "JETSCAPE",
    title = "{Multisystem Bayesian constraints on the transport coefficients of QCD matter}",
    eprint = "2011.01430",
    archivePrefix = "arXiv",
    primaryClass = "hep-ph",
    doi = "10.1103/PhysRevC.103.054904",
    journal = "Phys. Rev. C",
    volume = "103",
    number = "5",
    pages = "054904",
    year = "2021"
}

@article{JETSCAPE:2020shq,
    author = "Everett, D. and others",
    collaboration = "JETSCAPE",
    title = "{Phenomenological constraints on the transport properties of QCD matter with data-driven model averaging}",
    eprint = "2010.03928",
    archivePrefix = "arXiv",
    primaryClass = "hep-ph",
    doi = "10.1103/PhysRevLett.126.242301",
    journal = "Phys. Rev. Lett.",
    volume = "126",
    number = "24",
    pages = "242301",
    year = "2021"
}

@article{Nijs:2020roc,
    author = {Nijs, Govert and van der Schee, Wilke and G{\"u}rsoy, Umut and Snellings, Raimond},
    title = "{Bayesian analysis of heavy ion collisions with the heavy ion computational framework Trajectum}",
    eprint = "2010.15134",
    archivePrefix = "arXiv",
    primaryClass = "nucl-th",
    reportNumber = "CERN-TH-2020-175, MIT-CTP/5251",
    doi = "10.1103/PhysRevC.103.054909",
    journal = "Phys. Rev. C",
    volume = "103",
    number = "5",
    pages = "054909",
    year = "2021"
}

@article{Nijs:2020ors,
    author = {Nijs, Govert and van der Schee, Wilke and G{\"u}rsoy, Umut and Snellings, Raimond},
    title = "{Transverse Momentum Differential Global Analysis of Heavy-Ion Collisions}",
    eprint = "2010.15130",
    archivePrefix = "arXiv",
    primaryClass = "nucl-th",
    reportNumber = "CERN-TH-2020-174, MIT-CTP/5250",
    doi = "10.1103/PhysRevLett.126.202301",
    journal = "Phys. Rev. Lett.",
    volume = "126",
    number = "20",
    pages = "202301",
    year = "2021"
}

@article{Parkkila:2021tqq,
    author = "Parkkila, J. E. and Onnerstad, A. and Kim, D. J.",
    title = "{Bayesian estimation of the specific shear and bulk viscosity of the quark-gluon plasma with additional flow harmonic observables}",
    eprint = "2106.05019",
    archivePrefix = "arXiv",
    primaryClass = "hep-ph",
    doi = "10.1103/PhysRevC.104.054904",
    journal = "Phys. Rev. C",
    volume = "104",
    number = "5",
    pages = "054904",
    year = "2021"
}

@article{Mantysaari:2022ffw,
    author = {M{\"a}ntysaari, Heikki and Schenke, Bj{\"o}rn and Shen, Chun and Zhao, Wenbin},
    title = "{Bayesian inference of the fluctuating proton shape}",
    eprint = "2202.01998",
    archivePrefix = "arXiv",
    primaryClass = "hep-ph",
    doi = "10.1016/j.physletb.2022.137348",
    journal = "Phys. Lett. B",
    volume = "833",
    pages = "137348",
    year = "2022"
}

@article{Heffernan:2023gye,
    author = "Heffernan, Matthew R. and Gale, Charles and Jeon, Sangyong and Paquet, Jean-Fran{\c{c}}ois",
    title = "{Early-Times Yang-Mills Dynamics and the Characterization of Strongly Interacting Matter with Statistical Learning}",
    eprint = "2306.09619",
    archivePrefix = "arXiv",
    primaryClass = "nucl-th",
    reportNumber = "Phys. Rev. Lett. 132, 252301",
    doi = "10.1103/PhysRevLett.132.252301",
    journal = "Phys. Rev. Lett.",
    volume = "132",
    number = "25",
    pages = "252301",
    year = "2024"
}

@article{ALICE:2023tej,
    author = "Acharya, Shreyasi and others",
    collaboration = "ALICE",
    title = "{Skewness and kurtosis of mean transverse momentum fluctuations at the LHC energies}",
    eprint = "2308.16217",
    archivePrefix = "arXiv",
    primaryClass = "nucl-ex",
    reportNumber = "CERN-EP-2023-187",
    doi = "10.1016/j.physletb.2024.138541",
    journal = "Phys. Lett. B",
    volume = "850",
    pages = "138541",
    year = "2024"
}

@article{ALICE:2024apz,
    author = "Acharya, Shreyasi and others",
    collaboration = "ALICE",
    title = "{System size and energy dependence of the mean transverse momentum fluctuations at the LHC}",
    eprint = "2411.09334",
    archivePrefix = "arXiv",
    primaryClass = "nucl-ex",
    reportNumber = "CERN-EP-2024-297",
    doi = "10.1140/epjc/s10052-025-14325-4",
    journal = "Eur. Phys. J. C",
    volume = "85",
    number = "7",
    pages = "776",
    year = "2025"
}

@article{ATLAS:2024jvf,
    author = "Aad, Georges and others",
    collaboration = "ATLAS",
    title = "{Disentangling Sources of Momentum Fluctuations in Xe+Xe and Pb+Pb Collisions with the ATLAS Detector}",
    eprint = "2407.06413",
    archivePrefix = "arXiv",
    primaryClass = "nucl-ex",
    reportNumber = "CERN-EP-2024-173",
    doi = "10.1103/PhysRevLett.133.252301",
    journal = "Phys. Rev. Lett.",
    volume = "133",
    number = "25",
    pages = "252301",
    year = "2024"
}

@article{Saha:2025nyu,
    author = "Saha, Swati and Singh, Ranbir and Mohanty, Bedangadas",
    title = "{pT-differential radial flow in a blast-wave model}",
    eprint = "2505.19697",
    archivePrefix = "arXiv",
    primaryClass = "nucl-ex",
    doi = "10.1103/83zq-kdjg",
    journal = "Phys. Rev. C",
    volume = "112",
    number = "2",
    pages = "024902",
    year = "2025"
}

@article{Zhu:2026vql,
    author = "Zhu, Jie and Wu, Xiang-Yu and Qin, Guang-You",
    title = "{Mapping Nuclear Deformation with Differential Radial Flow in Heavy-Ion Collisions}",
    eprint = "2602.04148",
    archivePrefix = "arXiv",
    primaryClass = "nucl-th",
    month = "2",
    year = "2026"
}

@article{Bhatta:2025oyp,
    author = "Bhatta, Somadutta and Dimri, Aman and Jia, Jiangyong",
    title = "{Disentangling the global multiplicity and spectral shape fluctuations in radial flow}",
    eprint = "2504.20008",
    archivePrefix = "arXiv",
    primaryClass = "nucl-th",
    month = "4",
    year = "2025"
}

@article{Wan:2025rzg,
    author = "Wan, Jie and Wang, Chun-Zheng and Ma, Yu-Gang and Shou, Qi-Ye",
    title = "{Tracing pT-differential radial flow from blast-wave analytics to quark coalescence}",
    eprint = "2509.24889",
    archivePrefix = "arXiv",
    primaryClass = "nucl-th",
    month = "9",
    year = "2025"
}

@article{Sambataro:2025pop,
    author = "Sambataro, Maria Lucia and Plumari, Salvatore and Das, Santosh K. and Greco, Vincenzo",
    title = "{Probing the QGP through $p_T$-differential radial flow of heavy quarks}",
    eprint = "2510.19448",
    archivePrefix = "arXiv",
    primaryClass = "hep-ph",
    month = "10",
    year = "2025"
}

@article{CMS:2012zex,
    author = "Chatrchyan, Serguei and others",
    collaboration = "CMS",
    title = "{Measurement of the Elliptic Anisotropy of Charged Particles Produced in PbPb Collisions at $\sqrt{s}_{NN}$=2.76 TeV}",
    eprint = "1204.1409",
    archivePrefix = "arXiv",
    primaryClass = "nucl-ex",
    reportNumber = "CMS-HIN-10-002, CERN-PH-EP-2012-095",
    doi = "10.1103/PhysRevC.87.014902",
    journal = "Phys. Rev. C",
    volume = "87",
    number = "1",
    pages = "014902",
    year = "2013"
}

@article{CMS:2013bza,
    author = "Chatrchyan, Serguei and others",
    collaboration = "CMS",
    title = "{Studies of Azimuthal Dihadron Correlations in Ultra-Central PbPb Collisions at $\sqrt{s_{NN}} =$ 2.76 TeV}",
    eprint = "1312.1845",
    archivePrefix = "arXiv",
    primaryClass = "nucl-ex",
    reportNumber = "CMS-HIN-12-011, CERN-PH-EP-2013-214",
    doi = "10.1007/JHEP02(2014)088",
    journal = "JHEP",
    volume = "02",
    pages = "088",
    year = "2014"
}

@inproceedings{Hu:2026ebv,
    author = "Hu, Zixiao and McEwen, Jason D.",
    title = "{Efficient prior sensitivity analysis for Bayesian model comparison}",
    booktitle = "{44th International Workshop on Bayesian Inference and Maximum Entropy Methods in Science and Engineerin}",
    eprint = "2601.15132",
    archivePrefix = "arXiv",
    primaryClass = "stat.ME",
    month = "1",
    year = "2026"
}

@article{Han_2022,
   title={A Study on the Power Parameter in Power Prior Bayesian Analysis},
   volume={77},
   ISSN={1537-2731},
   url={http://dx.doi.org/10.1080/00031305.2022.2046160},
   DOI={10.1080/00031305.2022.2046160},
   number={1},
   journal={The American Statistician},
   publisher={Informa UK Limited},
   author={Han, Zifei and Ye, Keying and Wang, Min},
   year={2022},
   month=mar, pages={12–19} }

@article{Domingues:2026mis,
    author = "Domingues, Thiago S. and others",
    title = "{The shape of transverse momentum spectra in hybrid hydrodynamic models}",
    eprint = "2602.22490",
    archivePrefix = "arXiv",
    primaryClass = "nucl-th",
    month = "2",
    year = "2026"
}

@article{Parida:2026xbj,
    author = "Parida, Tribhuban and Samanta, Rupam and Ollitrault, Jean-Yves",
    title = "{The collectivity of transverse momentum fluctuations}",
    eprint = "2601.09890",
    archivePrefix = "arXiv",
    primaryClass = "nucl-th",
    doi = "10.1051/epjconf/202636405004",
    journal = "EPJ Web Conf.",
    volume = "364",
    pages = "05004",
    year = "2026"
}

@article{Agarwala:2025bxs,
    author = "Agarwala, Rohit and Basak, Dipankar and Dey, Kalyan",
    title = "{Observation of partonic collectivity via $p_{\rm T}$-differential radial flow fluctuations in Au+Au collisions at $\sqrt{s_{\rm NN}} = 200$ GeV}",
    eprint = "2512.15026",
    archivePrefix = "arXiv",
    primaryClass = "nucl-ex",
    month = "12",
    year = "2025"
}

@article{Du:2025hrz,
    author = "Du, Lipei and Jacobs, P. M.",
    title = "{Dynamical correlations across momentum scales in the Quark-Gluon Plasma}",
    eprint = "2512.10265",
    archivePrefix = "arXiv",
    primaryClass = "nucl-th",
    month = "12",
    year = "2025"
}

@article{Jia:2025rab,
    author = "Jia, Jiangyong",
    title = "{Sources of Radial-Flow Fluctuations in the Quark-Gluon Plasma}",
    eprint = "2507.14399",
    archivePrefix = "arXiv",
    primaryClass = "nucl-th",
    doi = "10.1103/d6mc-sq8m",
    journal = "Phys. Rev. Lett.",
    volume = "136",
    number = "11",
    pages = "112301",
    year = "2026"
}

@article{Jahan:2025uye,
    author = "Jahan, Syed Afrid and Roch, Hendrik and Shen, Chun",
    title = "{(3+1)D Bayesian analysis of RHIC BES data}",
    doi = "10.1016/j.jspc.2025.100209",
    journal = "J. Subatomic Part. Cosmol.",
    volume = "4",
    pages = "100209",
    year = "2025"
}

@article{Saltelli2002,
  author  = {Saltelli, A.},
  title   = {Making best use of model evaluations to compute sensitivity indices},
  journal = {Computer Physics Communications},
  volume  = {145},
  number  = {2},
  pages   = {280--297},
  year    = {2002},
  url     = {https://www.sciencedirect.com/science/article/abs/pii/S0010465502002801}
}

@article{Iwanaga2022,
  title = {Toward {SALib} 2.0: {Advancing} the accessibility and interpretability of global sensitivity analyses},
  volume = {4},
  url = {https://sesmo.org/article/view/18155},
  doi = {10.18174/sesmo.18155},
  journal = {Socio-Environmental Systems Modelling},
  author = {Iwanaga, Takuya and Usher, William and Herman, Jonathan},
  month = may,
  year = {2022},
  pages = {18155},
}

@article{Herman2017,
  doi = {10.21105/joss.00097},
  url = {https://doi.org/10.21105/joss.00097},
  year  = {2017},
  month = {jan},
  publisher = {The Open Journal},
  volume = {2},
  number = {9},
  author = {Jon Herman and Will Usher},
  title = {{SALib}: An open-source Python library for Sensitivity Analysis},
  journal = {The Journal of Open Source Software}
}

@article{SOBOL2001271,
title = {Global sensitivity indices for nonlinear mathematical models and their Monte Carlo estimates},
journal = {Mathematics and Computers in Simulation},
volume = {55},
number = {1},
pages = {271-280},
year = {2001},
note = {The Second IMACS Seminar on Monte Carlo Methods},
issn = {0378-4754},
doi = {https://doi.org/10.1016/S0378-4754(00)00270-6},
url = {https://www.sciencedirect.com/science/article/pii/S0378475400002706},
author = {I.M Sobol'}
}

@article{Auvinen:2026dyl,
    author = "Auvinen, Jussi and Eskola, Kari J. and Hirvonen, Henry and Niemi, Harri",
    title = "{Neural network enhanced Bayesian global analysis of relativistic heavy ion collisions}",
    eprint = "2603.26413",
    archivePrefix = "arXiv",
    primaryClass = "hep-ph",
    month = "3",
    year = "2026"
}

@article{scikit-learn,
  title={Scikit-learn: Machine Learning in {P}ython},
  author={Pedregosa, F. and Varoquaux, G. and Gramfort, A. and Michel, V.
          and Thirion, B. and Grisel, O. and Blondel, M. and Prettenhofer, P.
          and Weiss, R. and Dubourg, V. and Vanderplas, J. and Passos, A. and
          Cournapeau, D. and Brucher, M. and Perrot, M. and Duchesnay, E.},
  journal={Journal of Machine Learning Research},
  volume={12},
  pages={2825--2830},
  year={2011}
}

@article{Song:2009rh,
    author = "Song, Huichao and Heinz, Ulrich W",
    title = "{Interplay of shear and bulk viscosity in generating flow in heavy-ion collisions}",
    eprint = "0909.1549",
    archivePrefix = "arXiv",
    primaryClass = "nucl-th",
    doi = "10.1103/PhysRevC.81.024905",
    journal = "Phys. Rev. C",
    volume = "81",
    pages = "024905",
    year = "2010"
}

@article{Qian:2016fpi,
    author = "Qian, Jing and Heinz, Ulrich W. and Liu, Jia",
    title = "{Mode-coupling effects in anisotropic flow in heavy-ion collisions}",
    eprint = "1602.02813",
    archivePrefix = "arXiv",
    primaryClass = "nucl-th",
    doi = "10.1103/PhysRevC.93.064901",
    journal = "Phys. Rev. C",
    volume = "93",
    number = "6",
    pages = "064901",
    year = "2016"
}

@book{doi:10.1137/1.9781421407944,
author = {Golub, Gene H. and Van Loan, Charles F.},
title = {Matrix Computations - 4th Edition},
publisher = {Johns Hopkins University Press},
year = {2013},
doi = {10.1137/1.9781421407944},
address = {Philadelphia, PA},
edition   = {},
URL = {https://epubs.siam.org/doi/abs/10.1137/1.9781421407944},
eprint = {https://epubs.siam.org/doi/pdf/10.1137/1.9781421407944}
}

@article{harris2020numpy,
  title     = {Array programming with {NumPy}},
  author    = {Harris, Charles R. and Millman, K. Jarrod and van der Walt, St{\'e}fan J. and Gommers, Ralf and Virtanen, Pauli and Cournapeau, David and Wieser, Eric and Taylor, Julian and Berg, Sebastian and Smith, Nathaniel J. and Kern, Robert and Picus, Matti and Hoyer, Stephan and van Kerkwijk, Marten H. and Brett, Matthew and Haldane, Allan and del R{\'i}o, Jaime Fern{\'a}ndez and Wiebe, Mark and Peterson, Pearu and G{\'e}rard-Marchant, Pierre and Sheppard, Kevin and Reddy, Tyler and Weckesser, Warren and Abbasi, Hameer and Gohlke, Christoph and Oliphant, Travis E.},
  journal   = {Nature},
  volume    = {585},
  number    = {7825},
  pages     = {357--362},
  year      = {2020},
  publisher = {Springer Science and Business Media {LLC}},
  doi       = {10.1038/s41586-020-2649-2},
  url       = {https://doi.org}
}

@article{ALICE:2013wgn,
    author = "Abelev, Betty Bezverkhny and others",
    collaboration = "ALICE",
    title = "{Multiplicity Dependence of Pion, Kaon, Proton and Lambda Production in p-Pb Collisions at $\sqrt{s_{NN}}$ = 5.02 TeV}",
    eprint = "1307.6796",
    archivePrefix = "arXiv",
    primaryClass = "nucl-ex",
    reportNumber = "CERN-PH-EP-2013-135",
    doi = "10.1016/j.physletb.2013.11.020",
    journal = "Phys. Lett. B",
    volume = "728",
    pages = "25--38",
    year = "2014"
}

@techreport{CMS:2026oug,
      author = {{CMS Collaboration}},
      title         = "{Transverse-momentum differential radial flow fluctuations
                       across collision systems with CMS}",
      institution   = "CERN",
      reportNumber  = "CMS-PAS-HIN-25-016",
      address       = "Geneva",
      year          = "2026",
      url           = "https://cds.cern.ch/record/2965827",
}

@article{Wang:2026tvi,
    author = "Wang, Zaining",
    collaboration = "STAR",
    title = "{Collision energy and system size dependence of pT-differential radial flow fluctuations v0(pT) at RHIC}",
    eprint = "2608.00190",
    archivePrefix = "arXiv",
    primaryClass = "nucl-ex",
    doi = "10.1016/j.jspc.2026.100515",
    journal = "J. Subatomic Part. Cosmol.",
    volume = "6",
    pages = "100515",
    year = "2026"
}

@article{Thakur:2026few,
    author = "Thakur, Sanchari and Bhaduri, Partha Pratim",
    title = "{Characterizing $\pT$-differential Radial-Flow Fluctuations in Au+Au Collisions at FAIR with a Multi-Phase Transport Model}",
    eprint = "2610.41607",
    archivePrefix = "arXiv",
    primaryClass = "nucl-ex",
    month = "10",
    year = "2026"
}

@article{pileup,
   title="{Studies of azimuthal dihadron correlations in ultra-central PbPb collisions at $ \sqrt{{{s_{NN }
}}} $ =2.76 TeV}",
   volume={2014},
   ISSN={1029-8479},
   url={http://dx.doi.org/10.1007/JHEP02(2014)088},
   DOI={10.1007/jhep02(2014)088},
   number={2},
   journal={Journal of High Energy Physics},
   publisher={Springer Science and Business Media LLC},
   author={{CMS Collaboration}},
   year={2014},
   month=Feb }

@article{Lokhtin:2005px,
    author = "Lokhtin, I. P. and Snigirev, A. M.",
    title = "{A Model of jet quenching in ultrarelativistic heavy ion collisions and high-p(T) hadron spectra at RHIC}",
    eprint = "hep-ph/0506189",
    archivePrefix = "arXiv",
    doi = "10.1140/epjc/s2005-02426-3",
    journal = "Eur. Phys. J. C",
    volume = "45",
    pages = "211--217",
    year = "2006"
}

@article{AGOSTINELLI2003250,
    title = {Geant4---a simulation toolkit}, 
    author={{Geant4 Collaboration}},
	doi = {https://doi.org/10.1016/S0168-9002(03)01368-8},
	issn = {0168-9002},
	journal = {Nucl. Instrum. and Methods A},
	pages = {250-303},
	url = {https://www.sciencedirect.com/science/article/pii/S0168900203013688},
	volume = {506},
	year = {2003}
}

@article{Grad:1949zza,
    author = "Grad, Harold",
    title = "{On the kinetic theory of rarefied gases}",
    doi = "10.1002/cpa.3160020403",
    journal = "Commun. Pure Appl. Math.",
    volume = "2",
    number = "4",
    pages = "331--407",
    year = "1949"
}

@article{Dusling:2009df,
    author = "Dusling, Kevin and Moore, Guy D. and Teaney, Derek",
    title = "{Radiative energy loss and v(2) spectra for viscous hydrodynamics}",
    eprint = "0909.0754",
    archivePrefix = "arXiv",
    primaryClass = "nucl-th",
    doi = "10.1103/PhysRevC.81.034907",
    journal = "Phys. Rev. C",
    volume = "81",
    pages = "034907",
    year = "2010"
}

@article{Teaney:2003kp,
    author = "Teaney, Derek",
    title = "{The Effects of viscosity on spectra, elliptic flow, and HBT radii}",
    eprint = "nucl-th/0301099",
    archivePrefix = "arXiv",
    doi = "10.1103/PhysRevC.68.034913",
    journal = "Phys. Rev. C",
    volume = "68",
    pages = "034913",
    year = "2003"
}

@article{Monnai:2009ad,
    author = "Monnai, Akihiko and Hirano, Tetsufumi",
    title = "{Effects of Bulk Viscosity at Freezeout}",
    eprint = "0903.4436",
    archivePrefix = "arXiv",
    primaryClass = "nucl-th",
    doi = "10.1103/PhysRevC.80.054906",
    journal = "Phys. Rev. C",
    volume = "80",
    pages = "054906",
    year = "2009"
}

@article{Dusling:2011fd,
    author = {Dusling, Kevin and Sch{\"a}fer, Thomas},
    title = "{Bulk viscosity, particle spectra and flow in heavy-ion collisions}",
    eprint = "1109.5181",
    archivePrefix = "arXiv",
    primaryClass = "hep-ph",
    doi = "10.1103/PhysRevC.85.044909",
    journal = "Phys. Rev. C",
    volume = "85",
    pages = "044909",
    year = "2012"
}

@article{Denicol:2012cn,
    author = "Denicol, G. S. and Niemi, H. and Molnar, E. and Rischke, D. H.",
    title = "{Derivation of transient relativistic fluid dynamics from the Boltzmann equation}",
    eprint = "1202.4551",
    archivePrefix = "arXiv",
    primaryClass = "nucl-th",
    doi = "10.1103/PhysRevD.85.114047",
    journal = "Phys. Rev. D",
    volume = "85",
    pages = "114047",
    year = "2012",
    note = "[Erratum: Phys.Rev.D 91, 039902 (2015)]"
}

@article{Anderson:1974nyl,
    author = "Anderson, J. L. and Witting, H. R.",
    title = "{A relativistic relaxation-time model for the Boltzmann equation}",
    doi = "10.1016/0031-8914(74)90355-3",
    journal = "Physica",
    volume = "74",
    number = "3",
    pages = "466--488",
    year = "1974"
}

@article{Jaiswal:2014isa,
    author = "Jaiswal, Amaresh and Ryblewski, Radoslaw and Strickland, Michael",
    title = "{Transport coefficients for bulk viscous evolution in the relaxation time approximation}",
    eprint = "1407.7231",
    archivePrefix = "arXiv",
    primaryClass = "hep-ph",
    doi = "10.1103/PhysRevC.90.044908",
    journal = "Phys. Rev. C",
    volume = "90",
    number = "4",
    pages = "044908",
    year = "2014"
}

@article{Pratt:2010jt,
    author = "Pratt, Scott and Torrieri, Giorgio",
    title = "{Coupling Relativistic Viscous Hydrodynamics to Boltzmann Descriptions}",
    eprint = "1003.0413",
    archivePrefix = "arXiv",
    primaryClass = "nucl-th",
    doi = "10.1103/PhysRevC.82.044901",
    journal = "Phys. Rev. C",
    volume = "82",
    pages = "044901",
    year = "2010"
}

@article{McNelis:2019auj,
    author = "McNelis, Mike and Everett, Derek and Heinz, Ulrich",
    title = "{Particlization in fluid dynamical simulations of heavy-ion collisions: The i S3D module}",
    eprint = "1912.08271",
    archivePrefix = "arXiv",
    primaryClass = "nucl-th",
    doi = "10.1016/j.cpc.2020.107604",
    journal = "Comput. Phys. Commun.",
    volume = "258",
    pages = "107604",
    year = "2021"
}

@article{Bernhard:2016tnd,
    author = "Bernhard, Jonah E. and Moreland, J. Scott and Bass, Steffen A. and Liu, Jia and Heinz, Ulrich",
    title = "{Applying Bayesian parameter estimation to relativistic heavy-ion collisions: simultaneous characterization of the initial state and quark-gluon plasma medium}",
    eprint = "1605.03954",
    archivePrefix = "arXiv",
    primaryClass = "nucl-th",
    doi = "10.1103/PhysRevC.94.024907",
    journal = "Phys. Rev. C",
    volume = "94",
    number = "2",
    pages = "024907",
    year = "2016"
}
\end{document}